\documentclass[sigconf,screen]{acmart}

\setcopyright{rightsretained}
\acmPrice{}
\acmDOI{10.1145/3302516.3307357}
\acmYear{2019}
\copyrightyear{2019}
\acmISBN{978-1-4503-6277-1/19/02}
\acmConference[CC '19]{Proceedings of the 28th International Conference on Compiler Construction}{February 16--17, 2019}{Washington, DC, USA}
\acmBooktitle{Proceedings of the 28th International Conference on Compiler Construction (CC '19), February 16--17, 2019, Washington, DC, USA}

\usepackage{booktabs}   
\usepackage{subcaption} 

\usepackage[utf8]{inputenc}
\usepackage{tikz}
\usetikzlibrary{snakes,arrows,shapes}

\usepackage{graphicx}

\makeatletter
\newif\if@restonecol
\makeatother

\usepackage[ruled,vlined]{algorithm2e}

\usepackage{makecell}

\definecolor{xxxcolor}{rgb}{0.8,0,0}
\definecolor{maroon}{RGB}{128,0,64}
\definecolor{darkgreen}{RGB}{0,128,64}

\newcommand{\IGNORE}[1]{}

\newcommand{\tool}{\textsc{Revec}}

\usepackage{libertine}
\usepackage{fixltx2e}
\usepackage{xcolor}

\usepackage{enumitem}

\usepackage{balance}

\begin{document}

\title[\tool{}: Program Rejuvenation through Revectorization]{\tool{}: Program Rejuvenation through Revectorization}



\author{Charith Mendis}
\authornote{Both authors contributed equally to this research.}
\affiliation{
  \institution{MIT CSAIL}           
  \country{USA}                   
}
\email{charithm@mit.edu}         

\author{Ajay Jain}
\authornotemark[1]{}
\affiliation{
  \institution{MIT CSAIL}           
  \country{USA}                   
}
\email{ajayjain@mit.edu}

\author{Paras Jain}
\affiliation{
  \institution{UC Berkeley}           
  \country{USA}                   
}
\email{paras\_jain@berkeley.edu}         

\author{Saman Amarasinghe}
\affiliation{
  \institution{MIT CSAIL}           
  \country{USA}                   
}
\email{saman@csail.mit.edu}         

\begin{abstract}

  Modern microprocessors are equipped with Single Instruction Multiple Data (SIMD) or vector instructions which expose data level parallelism at a fine granularity. Programmers exploit this parallelism by using low-level vector intrinsics in their code. However, once programs are written using vector intrinsics of a specific instruction set, the code becomes non-portable. Modern compilers are unable to analyze and retarget the code to newer vector instruction sets. Hence, programmers have to manually rewrite the same code using vector intrinsics of a newer generation to exploit higher data widths and capabilities of new instruction sets. This process is tedious, error-prone and requires maintaining multiple code bases. We propose \tool{}, a compiler optimization pass which \emph{revectorizes} already vectorized code, by retargeting it to use vector instructions of newer generations. The transformation is transparent, happening at the compiler intermediate representation level, and enables performance portability of hand-vectorized code.

\tool{} can achieve performance improvements in real-world performance critical kernels. In particular, \tool{} achieves geometric mean speedups of 1.160$\times$ and 1.430$\times$ on fast integer unpacking kernels, and speedups of 1.145$\times$ and 1.195$\times$ on hand-vectorized x265 media codec kernels when retargeting their SSE-series implementations to use AVX2 and AVX-512 vector instructions respectively. We also extensively test \tool{}'s impact on 216 intrinsic-rich implementations of image processing and stencil kernels relative to hand-retargeting.

\end{abstract}



\maketitle


\section{Introduction}
\label{sec:intro}

Modern microprocessors have introduced SIMD or vector instruction sets to accelerate various performance critical applications by performing computations on multiple data items in parallel. Moreover, processor vendors have introduced multiple generations of vector instruction sets, each either increasing vector width or introducing newer computational capabilities. For example, Intel introduced MMX with 64-bit operations in 1997, Streaming SIMD Extensions (SSE) and 128 bit registers in 1999, SSE2, SSE3, SSSE3 and SSE4 from 2000--2006, AVX, AVX2 and FMA with 256 bit registers in 2011, and AVX-512 and 512 bit registers in 2016.
SIMD instruction sets from other processor vendors include AMD's 3DNow!~\cite{amd3dnow}, IBM's VMX/Altivec~\cite{vmx} and ARM's Neon~\cite{neon}. In order to use these SIMD units, programmers must either hand-code directly using platform-specific intrinsics, or rely on existing compiler auto-vectorization techniques to discover opportunities in programs written in mid- or high-level languages.

Modern compilers employ two main auto-vectorization strategies, namely, loop vectorization~\cite{FortranLoop} and Superword Level Parallelism (SLP) based vectorization~\cite{LarsenSLP}. Auto-vec\-torization allows programmers to write code in high-level languages, while still benefiting from SIMD code generation.
However, both loop and SLP vectorization rely on programmers writing code in ways which expose existing data level parallelism.
In certain cases, the programmer needs to know the underlying implementation of compiler vectorization passes to cater her code writing style. Even then, auto-vectorization may not vectorize all vectorizable code regions due to inaccuracies in cost models, inability to perform certain transformations etc~\cite{maleki2011evaluation}.

In contrast, writing hand-vectorized code allows the programmers to exploit fine-grained parallelism in the
programs more precisely. Hand-vectorization allows programmers to explicitly embed domain optimizations
such as performing intermediate operations without type promotions,
which cannot be achieved by compiler auto-vectorization. However, when manual vectorizing,
programmers give up on both code and performance portability.
The code that is the fastest for an older vector instruction set may perform suboptimally on a processor which supports wider vector instructions. For example, manually vectorized SSE2 code will not utilize the full data width of a processor that supports AVX2 instructions. This issue is aggravated as modern compilers do not retarget code written in low-level intrinsics to use newer vector instructions. Hence, programmers frequently maintain several architecture-specific, version-specific implementations of each computationally intensive routine in a codebase to exploit newer instructions.
This is tedious, error-prone, and is a maintenance burden.

In this work, we propose compiler \emph{revectorization}, the retargeting of hand-vectorized code to use newer vector instructions of higher vector width. We developed \tool{}, a compiler optimization technique to algorithmically achieve revectorization, and implemented it in the LLVM compiler infrastructure~\cite{LLVM} as an IR level pass. \tool{} rejuvenates performance of stale implementations of data-parallel portions of hand vectorized programs, automatically adding performance portability.

\tool{} finds opportunities to merge two or more similar vector instructions to form vector instructions of higher width. \tool{} has its foundations in SLP auto-vectorization~\cite{LarsenSLP}, but rather than transforming scalar code, focuses only on revectorizing already vectorized code and hence has its own unique challenges. More specifically, \tool{} needs to find equivalences between vector intrinsics with complex semantics and decide how to merge and retarget vector shuffle instructions to newer instruction sets.

\tool{} automatically finds equivalences between vector intrinsics across different instruction
generations by enumerating all combinations. Equivalences are established through randomized
and corner case testing. \tool{} also introduces vector shuffle merge patterns to handle revectorizing shuffle instructions.
During compiler transformation, \tool{} first does loop unrolling and reduction variable splitting,
with heuristics that are catered towards revectorization. These preprocessing transformations
expose more opportunities for revectorization. Finally, it uses the automatically found equivalences
and shuffle merge rules to merge two or more vector instructions of a lower data width to form
vector instructions of a higher data width.
\subsection{Contributions}
In this paper, we make the following contributions:
\begin{itemize}[leftmargin=0.5cm,topsep=3pt]
\item A compiler optimization technique, \tool{}, which automatically converts the hand-vectorized code to vector instructions of higher vector width including computations that involve reductions.
\item Automatically finding instances of two or more vector intrinsics of similar semantics which can be merged into a vector instruction of higher width by enumerating and refining candidates.
\item Vector shuffle merge rules to enable revectorizing vector shuffle instructions.
\item Implementation of \tool{} in LLVM compiler infrastructure as an LLVM IR level optimization pass to transparently perform revectorization.
\item Extensive evaluation of \tool{} on real-world performance critical kernels ranging from media compression codecs, integer compression schemes, image processing kernels and stencil kernels. We show \tool{} automatically achieves geometric mean speedups of 1.145$\times$ and 1.195$\times$ on hand-vectorized x265 media codec kernels, and speedups of 1.160$\times$ and 1.430$\times$ on fast integer unpacking kernels when retargeting their SSE implementations to use AVX2 and AVX512 vector instructions respectively. Further, \tool{} achieves geometric mean speedups of 1.102$\times$ and 1.116$\times$ automatically on 216 intrinsic-rich implementations of image processing and stencil kernels written using SSE-series (SSE2+) vector instructions.
\end{itemize}

\section{Motivation}
\label{section:motivation}

We use the \texttt{MeanFilter3x3} kernel from the Simd Image processing library~\cite{simd} to motivate how \tool{} achieves performance portability of hand-vectorized code. Consider different \texttt{MeanFilter3x3}
\begin{figure*}[t]
  \includegraphics[width=\textwidth]{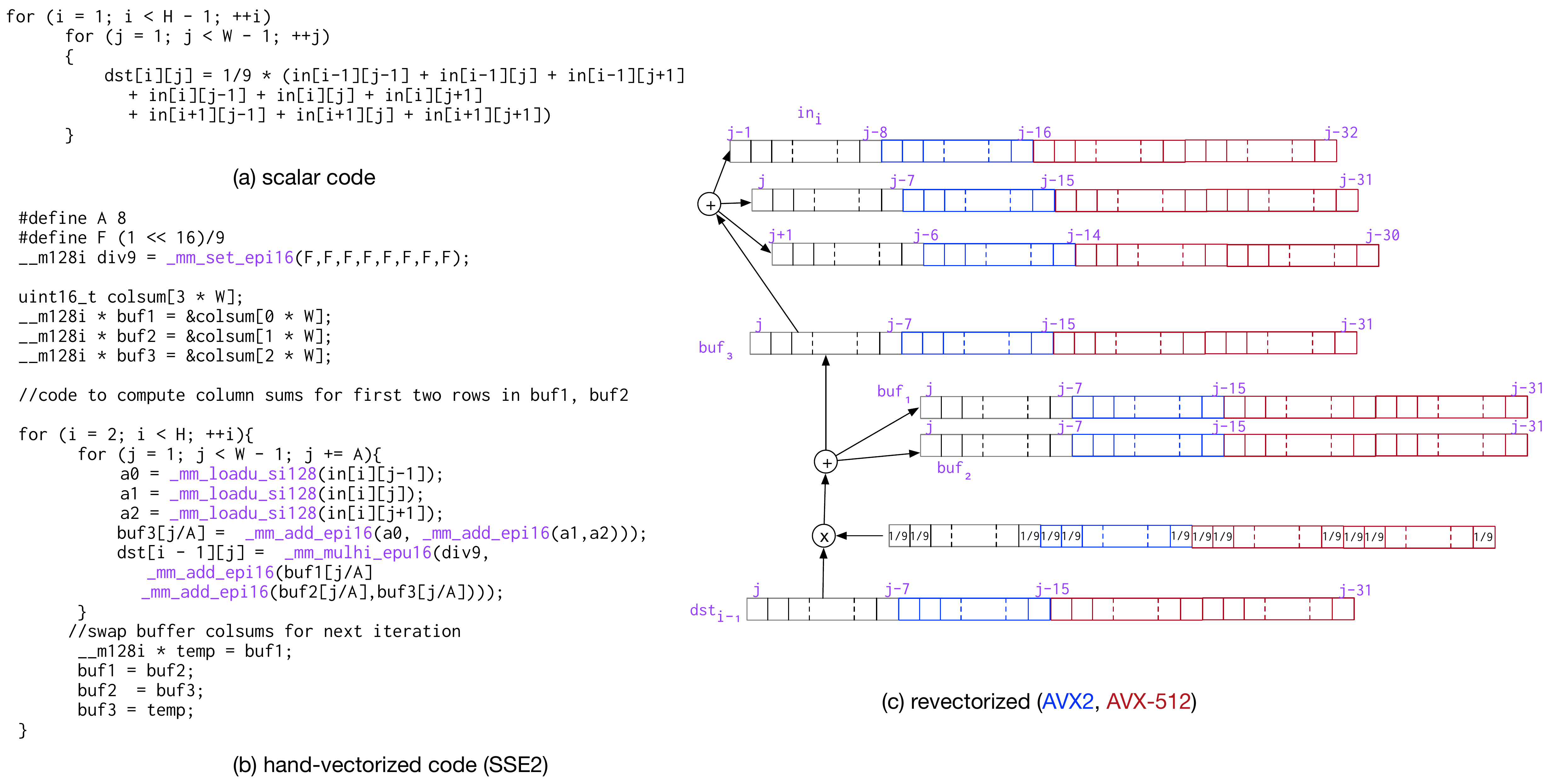}
  \caption{(a) scalar code and (b) simplified hand-vectorized SSE2 code for \texttt{Meanfilter3x3} from Simd image processing library. (c) Computation Graph for the revectorized versions targeting AVX2 (256 bits) and AVX-512 (512 bits) for the inner loop of SSE2 version. Note that handling of boundary conditions and type conversions are omitted for clarity of presentation.}
 \label{fig:motivation}
\end{figure*}
implementations shown in Figure~\ref{fig:motivation} using C-like code. Pixel values of an output image are computed by averaging the corresponding pixel values in a 3 by 3 window of an input image.
Figure~\ref{fig:motivation}(a) and (b) show the scalar version and the SSE2 hand-vectorized version of the mean filter respectively.
Note that boundary condition handling and certain type conversions from the original implementations are omitted for the sake of clarity.
\begin{figure*}[t]
  \includegraphics[width=0.9\textwidth]{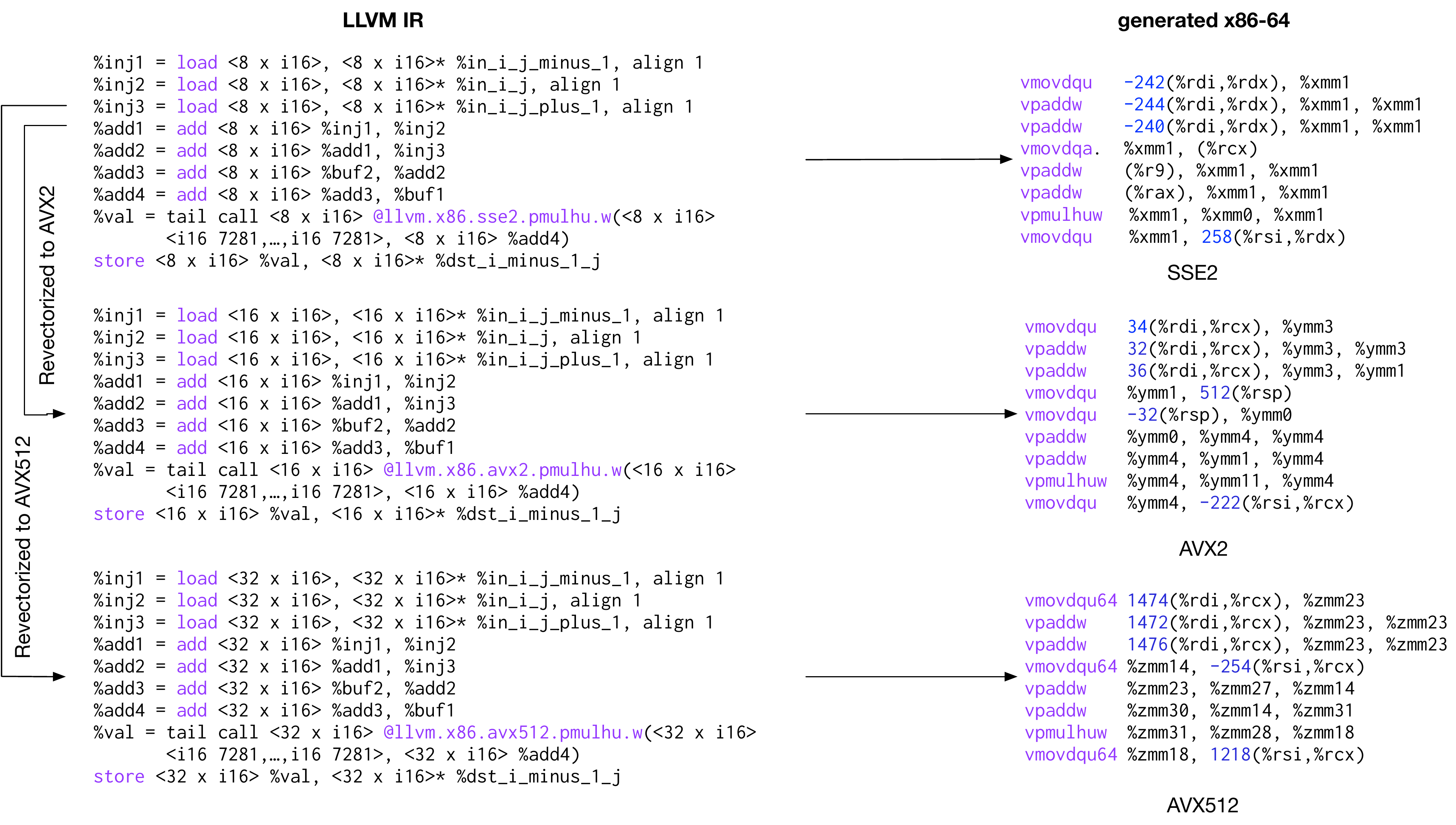}
  \caption{LLVM IR and x86-64 assembly code snippets for the inner loop of original SSE2 implementation of \texttt{MeanFilter3x3} and revectorized output targetting AVX2 and AVX-512 instruction sets. }
 \label{fig:revec}
\end{figure*}
The scalar version implements a naive algorithm which goes through each output pixel and individually computes the mean of the corresponding pixels
in the input image. It is computationally inefficient since it recomputes column sums for each row repeatedly.
However, LLVM is able to auto-vectorize this scalar kernel without much analysis, and hence its performance scales with increases in the width of vector instructions.

The SSE2 version is algorithmically different from the scalar version. It maintains an array to store column sums of 3 adjacent rows.
The inner loop computes the column sums of the bottom-most row (of the 3 rows) 8 elements at a time using SSE2 vector instructions.
The output image pixel values are computed, again 8 elements at a time by
summing up the column sums for the 3 adjacent rows and finally by multiplying it with 1/9. The algorithm used in this version is computationally
efficient and is explicitly vectorized to exploit data level parallelism.

Algorithms used in~\ref{fig:motivation}(a) and (b) are both data parallel. However, transforming (a) to (b) automatically is non-trivial.
Further, we found that LLVM fails to auto-vectorize the scalarized, buffer version of~\ref{fig:motivation}(b). This shows that considerable human insight goes into developing
work efficient hand-vectorized data parallel algorithms while compiler auto-vec\-torization is not always reliable.

However, once code is written using vector intrinsics, compilers skip analyzing vector code to enable further vectorization.
The assembly code generated for the SSE2 implementation when targeted to a Skylake processor with AVX-512 extensions still produces only SSE2 code (Figure~\ref{fig:revec}).
Even though the SSE2 version is fast, it cannot fully utilize the wider vector instructions available in modern processors with AVX2, AVX-512 support.
Essentially, hand-vectorized code loses performance portability, and programmers have to rewrite the same code using newer vector instructions to exploit wider vector
instructions.

\tool{} adds performance portability to hand-vectorized code by automatically retargeting it to vector instructions of higher vector width,
whenever available. Figure~\ref{fig:motivation}(c) shows the computation graphs of widened instructions by \tool{} for the inner loop of the SSE2 version.
It uses vector intrinsic equivalences found across vector instruction sets and vector shuffle rules to widen the vector width of already vectorized code by
merging two or more narrow vector width instructions akin to SLP vectorization~\cite{LarsenSLP}. For example, to form AVX2 and AVX-512 instructions for the \texttt{MeanFilter3x3}
example, two and four SSE2 instructions are merged respectively. Given the SSE2 implementation of \texttt{MeanFilter3x3} found in the Simd library, \tool{} achieves a 1.304$\times$ speedup when retargeting to AVX-512.

\tool{} is developed as an LLVM IR level transformation pass and does revectorization transparently without human involvement. Figure~\ref{fig:revec} shows the LLVM IR
and corresponding x86-64 assembly instructions when targeting AVX2 and AVX-512 instruction sets under \tool{}.
In comparison,
stock LLVM compiler which does not revectorize vectorized code generates only SSE2 assembly instructions. We also compiled the SSE2 implementation under GCC5.4 and ICC and found that
none of the compilers were able to revectorize the code, with similar limitations to LLVM.

\begin{figure}[t]
  \includegraphics[width=\columnwidth]{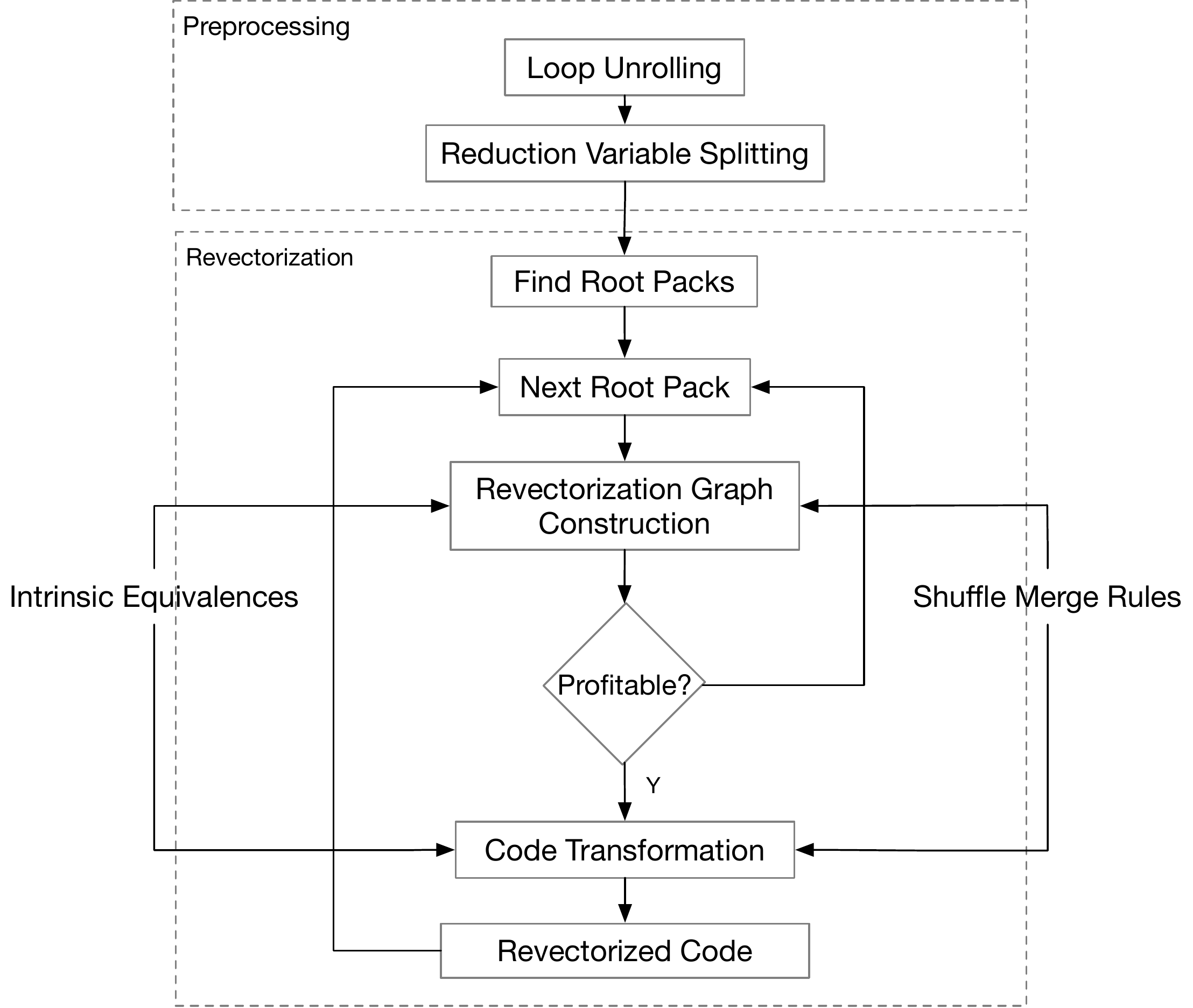}
  \caption{Overview of the \tool{} compiler transformation pass.}
 \label{fig:overview}
\end{figure}

\section{\tool{} Overview}
\label{sec:architecture}

Figure~\ref{fig:overview} shows the high level overview of \tool{} compiler transformation pass.
It first performs preprocessing transformations -- loop unrolling and reduction variable splitting --
which expose opportunities for revectorization (Section~\ref{sec:preprocess}).
\tool{} then transforms the code to use vector instructions of higher width using a technique akin to bottom-up SLP vectorization~\cite{Rosen2007}, but specialized to handle vector instructions.

\tool{} first finds adjacent vector stores and vector $\phi$-node instructions~\cite{ssa-form}
of same data type in a basic block.
These act as the initial \emph{revectorization packs}, which are bundles of vector instructions
which can be merged into a single vector instruction of higher vector width. Starting from the
initial packs which act as roots, \tool{} next builds a \emph{revectorization graph}
following their use-def chains (Section~\ref{sec:graph}).
\tool{} adds packs to the graph recursively until it terminates with vector
instructions which are not mergeable. \tool{} uses equivalences found among vector
intrinsics (Section~\ref{sec:enumeration}) and vector shuffle rules
(Section~\ref{sec:shuffle}) to determine whether two intrinsics can be merged.

Next, \tool{} determines the
profitability of the revectorization graph using a cost model. If revectorization is profitable, the graph is transformed to use vector instructions of higher width. \tool{} continues to iteratively revectorize until all roots are exhausted for a given basic block.

\paragraph{Notation} We implement \tool{} as a middle end compiler pass in the LLVM compiler infrastructure to transform LLVM IR instructions. In subsequent sections, we use the LLVM IR vector notation to explain the functionality of \tool{}. LLVM vector types are represented as \texttt{v = <m x ty>}, where vector type \texttt{v} contains \texttt{m} elements of scalar type \texttt{ty}. A generic LLVM IR vector instruction has the following form.

\begin{center}
  \texttt{out = op <m x ty> opnd1, opnd2, ..., opndn}
\end{center}

\texttt{op} is the opcode of the instruction, \texttt{opnd1} up to \texttt{opndn} are operands and \texttt{out} is the output value of the instruction. Note that output type is made explicit in the instruction, whereas operand types are omitted. Operand types are made explicit in instances when it is not clear from the context.

We call vector instructions which do not use the complete data width of the processor as \emph{narrow vectors}, and these are widened through revectorization. We denote a revectorization pack consisting of $I_1, I_2, ..., I_n$ vector instructions by $\{I_1, I_2, ..., I_n\}$. Also we define \emph{vectorization factor} to be the number of vectors packed together in a revectorization pack.

\paragraph{Running Example} We use the example shown in Figure~\ref{fig:example} to illustrate \tool{}'s functionality when targeting a processor with AVX2 instructions. The code shown in Figure~\ref{fig:example}(a) widens the data of the \texttt{in} array from 16 bits to 32 bits using vector shuffle instructions and then adds a constant to each element. Finally, the added values are downcasted back to 16 bits and are stored in the \texttt{out} array. All operations are written using SSE2 vector intrinsics.

\section{Preprocessing Transformations}
\label{sec:preprocess}

\tool{} performs two preprocessing transformations to expose opportunities for revectorization:
\emph{Loop Unrolling} and \emph{Reduction Variable Splitting}.

\subsection{Loop Unrolling}

\tool{} first unrolls inner loops containing vector instructions to expose opportunities for merging narrow vector instructions.
First, \tool{} traverses each basic block within a given inner loop and finds all vectorized stores. These stores are next separated into mulitple store chains, where each chain contains a set of vectorized stores which access adjacent memory locations. \tool{} considers each store chain, beginning with the chain with the fewest number of stores, to update the unroll factor for the inner loop.

If the cumulative width of a given store chain (SCW = size of a store $\times$ number of stores) is less than the maximum vector width of the processor (VW), then there is an opportunity for revectorization. For such chains, \tool{} checks if unrolling leads to a longer chain of consecutive stores by checking if the symbolic address of the tail of the store chain plus the size of a store is equal to the symbolic address of the head of the store chain in the next iteration.
Symbolic address information is available through the Scalar Evolution pass of LLVM.
If this condition is true, unrolling leads to a longer store chain. Then, the Unroll Factor (UF) is updated:
\begin{equation}
  \label{eq:unroll}
  \text{UF} = \max{(\frac{\texttt{lcm}(\text{VW},\text{SCW})}{\text{SCW}},\text{UF})}
\end{equation}
\tool{} initializes UF = 1 and iterates through all store chains in
basic blocks of an inner loop to update its unroll factor.

Next, \tool{} checks if there are any vector
$\phi$-node instructions that are part of a reduction chain. $\phi$-nodes are used in
Single Static Assignment~\cite{ssa-form}
based IRs to select a value based on the predecessor of the current basic block. If any such $\phi$-nodes are found,
\tool{} updates the unroll factor analogous to equation~\ref{eq:unroll}, replacing SCW with the width of the $\phi$-node instruction.

\begin{figure*}[h]
  \includegraphics[width=\textwidth]{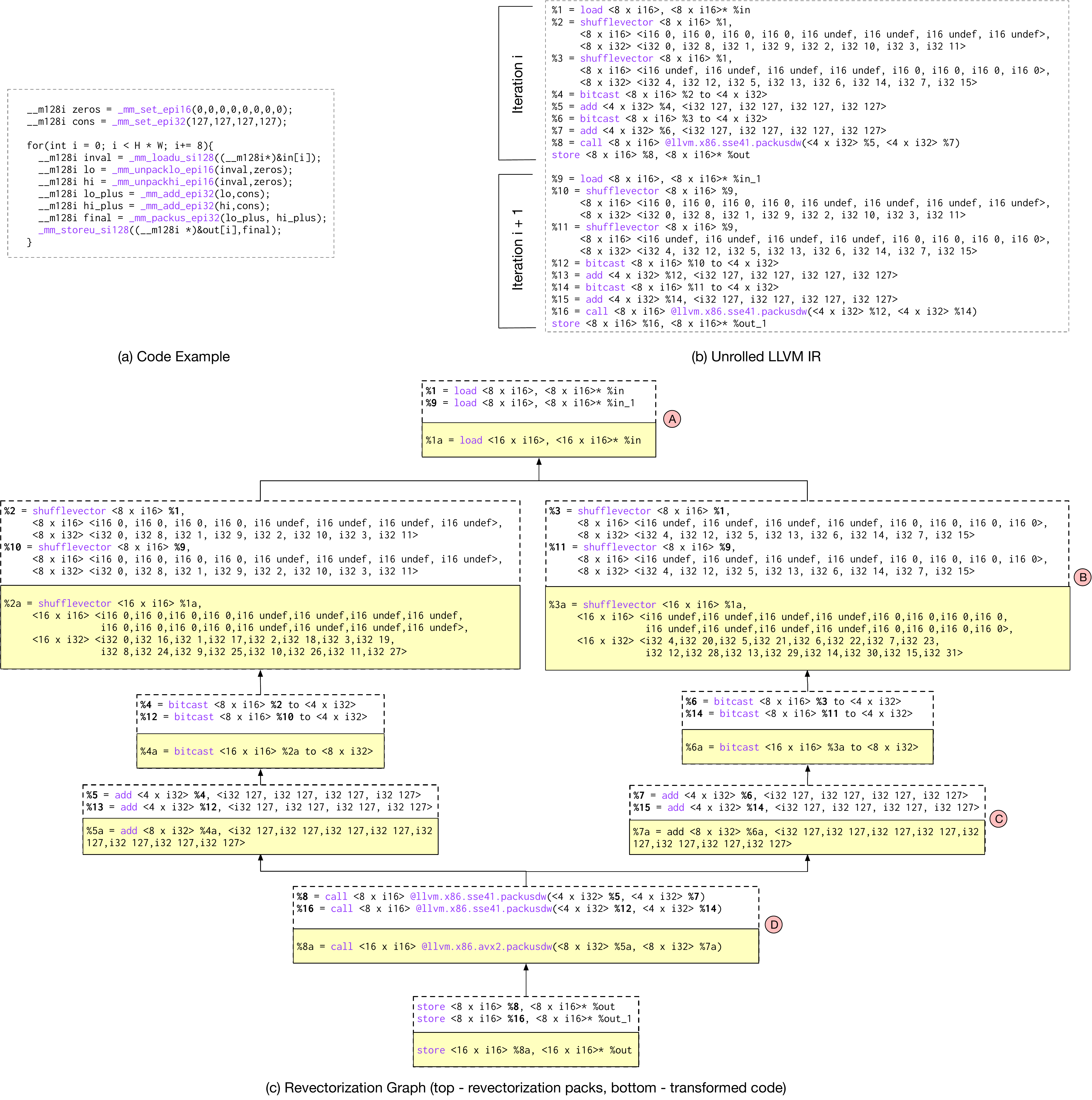}
    \caption{(a) Code example written using SSE2 vector intrinsics (b) LLVM IR after unrolling on a server with AVX2 instructions (c) Revectorization graph for the IR where the top half of the node shows the revectorization pack associated with that node and the bottom half shows the transformed IR after revectorization targeting AVX2.}
 \label{fig:example}
\end{figure*}

Finally, \tool{} unrolls the loop using this unroll factor. Figure~\ref{fig:example}(b)
shows the unrolled LLVM IR for the code shown in Figure~\ref{fig:example}(a). Since there is
only one 128 bit store within the loop, the unroll factor for an AVX2 processor is 2.

\subsection{Reduction Variable Splitting}
\label{sec:red-split}

\begin{figure*}[ht]
  \includegraphics[width=\textwidth]{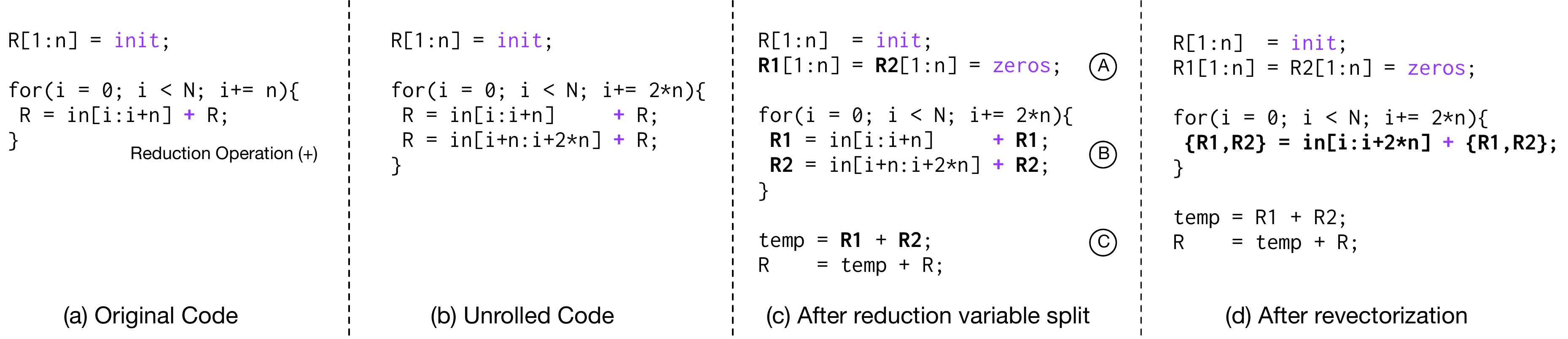}
  \caption{(a) Code example with a reduction (b) Code after the loop is unrolled twice
    (c) Code after reduction variable \texttt{R} is splitted into \texttt{R1} and \texttt{R2}
    (d) New independent reduction variables \texttt{R1} and \texttt{R2} are revectorized.}
  \label{fig:reduction}
\end{figure*}

Reduction computations involve accumulation of a set of values
into one reduction variable using a reduction operation. For example,
consider the code snippet in Figure~\ref{fig:reduction}(a), where values of
array \texttt{in} are added together into the reduction variable \texttt{R}.
Reductions are not explicitly parallelizable and hence are not explicitly revectorizable.
However, provided that the reduction operation is associative, we could use multiple
accumulators in parallel to perform reduction of different parts of the data and finally
accumulate the results into the original variable.
To achieve this data partitioning, we introduce multiple independent
reduction variables which accumulate results of different parts of the data making it
amenable to revectorization.

Concretely, after loop unrolling, \tool{} first identifies reduction variables
with associative operations such as maximum, addition etc. whose values
are only consumed outside the loop in which the reduction is computed. Next, for each update
of the reduction, \tool{} introduces a new reduction variable by \emph{splitting}
the original variable. For example, in Figure~\ref{fig:reduction}(c)\textcircled{\small B},
\tool{} introduces two independent reduction variables, \texttt{R1} and \texttt{R2}, in place of the
two occurences of \texttt{R} within the loop. Finally, \tool{} emits cleanup prologue
\textcircled{\small A} and epilogue \textcircled{\small C} code outside the loop,
to first initialize the reduction variables and then to do the final extraction and
accumulation of values of newly introduced
reduction variables into the original reduction variable. Now, the reduction computation inside the loop
is revectorizable. Figure~\ref{fig:reduction} shows the final revectorized reduction code, where \texttt{R1}
and \texttt{R2} values are computed in parallel.

\section{Revectorization Graph Construction}
\label{sec:graph}

\tool{} first finds the initial revectorization packs which form the roots of the revectorization graph. Then, \tool{} recursively builds a graph of packs by following the use-def chains of the instructions at the frontier of the graph.

\subsection{Finding Root Packs}
\label{ssec:finding_roots}

For each basic block, \tool{} first finds chains of vector store instructions with adjacent
memory accesses and chains of $\phi$-node vector instructions of same data type.
These two types of instructions are generally part of use-def chains with high computational
workload and hence are suitable candidates to seed revectorization.
Store instructions act as terminal instructions which write back values to memory after computation and 
$\phi$-node instructions act as entry points of a basic block with reductions.
\tool{} forms initial revectorization packs from these chains.

For a target with maximum vector width $VW$,
and a chain of vector seeds with operand width $SW$, the maximum
number of vector instructions that can be packed together is
$VF_{\text{max}} = \frac{VW}{SW}$. \tool{} takes chunks of $VF_{\text{max}}$
instructions from each vector chain and forms the
initial revectorization packs. These packs act as the roots of their
respective revectorization graphs. For example, the two stores in the LLVM IR
shown in Figure~\ref{fig:example}(b) are packed to form the root of the
revectorization graph in Figure~\ref{fig:example}(c).

\subsection{Building the Graph}

\tool{} grows the graph by following the use-def chains of packs at the frontier of the graph similar to bottom-up SLP~\cite{Rosen2007}. \tool{} iterates through all operands of a pack in order. If their definitions can be merged into vectors of higher width, \tool{} adds the packs for the operands as new nodes of the graph with an edge connecting the current pack and the operand pack. The graph is built recursively traversing the operands of already formed packs. Expansion stops when all packs in the frontier have vector instructions which are not mergeable. Two or more instructions in a revectorization pack can be merged if the following conditions are met:

\begin{itemize}[leftmargin=0.5cm]
\item Isomorphic: They perform the same operation on operands of the same data type and return vector values of the same data type, if any. \tool{} handles both vector intrinsics which
  are lifted to LLVM vector IR form (e.g., Figure~\ref{fig:example}(c) Pack \textcircled{\small C}) and vector intrinsics which remain as opaque LLVM intrinsic functions (e.g., Figure~\ref{fig:example}(c) Pack \textcircled{\small D}). \IGNORE{Equivalences and transformation rules for the later are found via enumeration (Section~\ref{sec:enumeration}).} 
\item Independent: Their operands should be independent of the results they produce.
\item Adjacent Memory Accesses: If instructions access memory, their accesses should be adjacent (e.g., Figure~\ref{fig:example}(c) Pack \textcircled{\small A}).
\item Vector Shuffles: Special shuffle rules (Section~\ref{sec:shuffle}) are applied to merge vector shuffles and to grow the graph beyond shuffles (e.g., Figure~\ref{fig:example}(c) Pack \textcircled{\small B}).
\end{itemize}

In general, a revectorization graph is a Directed Acyclic Graph (DAG), where a single pack can be used by multiple other packs. Figure~\ref{fig:example}(c) shows the complete revectorization graph for our running example. Note that the packs are shown in the upper half of the nodes, whereas the bottom half shows the revectorized LLVM IR instruction after code transformation (Section~\ref{sec:transform}).

\section{Code Transformation}
\label{sec:transform}

Given a revectorization graph, \tool{} first decides whether it is profitable to transform the code to use wider vectors using a static cost model.
Finally, \tool{} revectorizes graphs which are deemed profitable.

\subsection{Profitability Analysis}
\label{subsec:cost}

With an additive cost model, \tool{} sums the benefit of revectorizing each pack in a graph to the find the cumulative benefit of the whole strategy.
Replacing multiple narrow vector instructions with a single instruction is typically profitable, evaluated by querying the LLVM \texttt{TargetTransform\-Info} interface. However, gathering non\-revectorizable packs or non-adjacent loads at the leaves of the revectorization graph has overhead. Further, if a pack contains narrow vectors that are used out-of-graph, \tool{} must emit a subvector extract instruction. For a particular graph, \tool{} only transforms the code if the total benefit of revectorizing packs is greater than the total gather and extract cost.

\subsection{Revectorizing the Graph}

Transformation happens recursively while traversing the graph starting from the roots.
\tool{} invokes the RevectorizeGraph routine (Algorithm~\ref{alg:revec}) with initial revectorization packs at the
roots of the graph.

If the current pack is not mergeable to a wider vector instruction, \tool{} gathers its constituents using a tree of vector
shuffle instructions (Section~\ref{sec:gather}). If it is revectorizable, \tool{} handles revectorizing vector shuffle instructions
as detailed in Section~\ref{sec:shuffle}. For other vector instructions, \tool{} uses a generic widening strategy (Section~\ref{sec:wide}) covering
both vector instructions which are lifted to LLVM vector IR form and instructions which remain opaque LLVM vector intrinsics. The tree is traversed depth first
to generate revectorized values of children (operands) first and these values are used to generate the revectorized values for the current pack.
\texttt{IntrinsicMap} is an auto-generated map used for widening opaque LLVM vector intrinsics, which maps from narrow vector intrinsics to wider vector
intrinsics at various vectorization factors. Generation of \texttt{IntrinsicMap} is detailed in Section~\ref{sec:enumeration}.

\begin{algorithm}[t]
  \small
 \eIf{!IsMergeable(Pack)}{
  \Return TreeGather(Pack) \tcp*{Section \ref{sec:gather}}
 }{

   \If(\tcp*[f]{Section \ref{sec:shuffle}}){ContainsShuffles(Pack)}{
     \Return RevectorizeShuffle(Pack)
   }

  V = $\phi$ \\
  \For{ChPack $\in$ Children(Pack)}{
    V.add(RevectorizeGraph(ChPack, IntrinsicMap))
  }

\eIf(\tcp*[f]{Section \ref{sec:wide}}){ContainsIntrinsics(Pack)}{
    \Return RevectorizeIntrinsic(Pack, V, IntrinsicMap)
  }{
    \Return RevectorizeVectorIR(Pack, V)
  }
 }
 \caption{RevectorizeGraph(Pack, IntrinsicMap)}
 \label{alg:revec}
\end{algorithm}

\subsection{Generic Widening of Vectors}
\label{sec:wide}

\tool{} widens both packs of vector intrinsics that have been lifted to high level IR and packs of vector intrinsics which remain as opaque calls.
For instance, LLVM lifts the intrinsic \texttt{\_mm\_add\_epi16} to the vector IR instruction \texttt{add <8 x i16>}, as the semantics of the intrinsic and IR instruction have a known correspondence. However, many intrinsics have complex semantics that do not map to simple compiler IR. For example, LLVM propagates the intrinsic \texttt{\_mm\_packus\_epi32} to the backend with the LLVM IR intrinsic \texttt{@llvm.x86.sse41.packusdw}.

\paragraph{Widening Lifted IR} If the $i$-th instruction in a pack of $p$ isomorphic vector IR instructions has the generic form,
$$\texttt{out}_\texttt{i} = \texttt{op <m x ty>} \quad \texttt{opnd}_{\texttt{i1}},~\texttt{opnd}_{\texttt{i2}},~...,~\texttt{opnd}_{\texttt{ik}}$$

the widened instruction becomes, $$\texttt{out} = \texttt{op <(p*m) x ty>} \quad \texttt{V}_\texttt{1},~\texttt{V}_\texttt{2},~..., \texttt{V}_\texttt{p}$$ where $V_1,V_2,...,V_p$ are revectorized values of the operand packs (children of the original pack in the revectorization graph). Note that the opcode is reused. Operands in different lanes need not have the same type. However, as we build packs recursively from vector operands, only instructions that yield vectors are widened. Scalar packing is more appropriate for the scalar SLP autovectorizer.

\paragraph{Widening Intrinsic Calls} For a pack of isomorphic intrinsic calls, \tool{} queries \texttt{IntrinsicMap} to find a wider vector intrinsic. \tool{} then transforms the pack by first packing the operands as before, then emitting a single intrinsic call to the widened conversion found. Figure \ref{fig:example}(c) pack \textcircled{\small D} illustrates opcode substitutions for the LLVM intrinsic \texttt{@llvm.x86.sse41.packusdw}.

\subsection{Transforming Packs of Vector Shuffles}
\label{sec:shuffle}

A vector shuffle is a high level instruction that reorders scalars from lanes of two vector operands based on a constant indexing mask. A shuffle generalizes vector selects, permutations with arbitrary masks, operand interleaving, and two-operand gathers. In the shuffle $\texttt{s = shuffle a, b, m}$, let \texttt{a} and \texttt{b} be source vector operands of the same type, \texttt{m} be a length $l$ constant vector,
and the shuffled value $\texttt{s}$ be a length $l$ vector created by indexing into the concatenation of \texttt{a} and \texttt{b} at each index in $m$.
During code transformation, \tool{} matches shuffle packs to known operand patterns to create efficient revectorized shuffles.

In general, consider a pack of vector shuffles:
\begin{align*}
    \texttt{s\textsubscript{1} } &\texttt{= shuffle a\textsubscript{1}, b\textsubscript{1}, m\textsubscript{1}} ~...\\
    \texttt{s\textsubscript{p} } &\texttt{= shuffle a\textsubscript{p}, b\textsubscript{p}, m\textsubscript{p}}
\end{align*}

\begin{figure}[t]
    \centering
    \includegraphics[width=\linewidth]{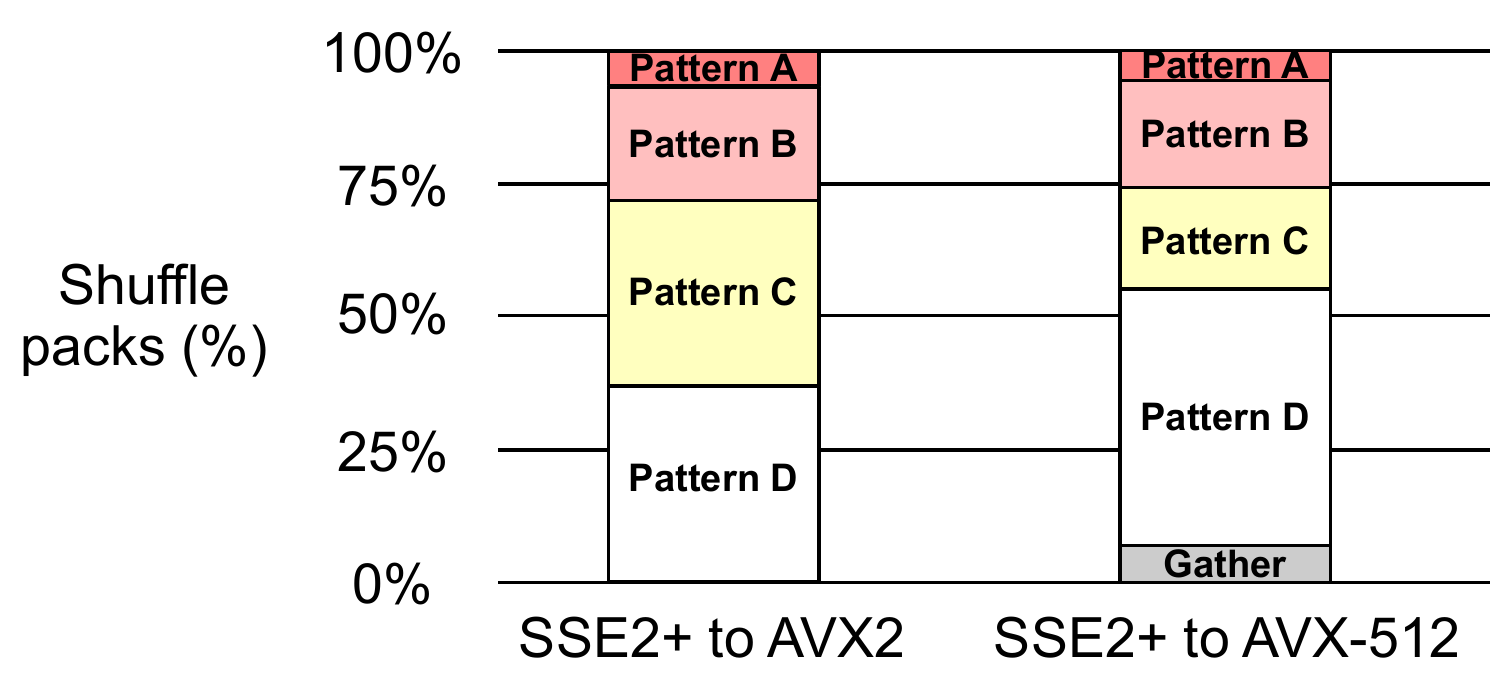}
    \caption{\tool{}'s shuffle patterns match all shuffle packs encountered in benchmarks when revectorizing to AVX2, and 92.9\% of shuffle packs encountered when revectorizing to AVX-512.}
    \label{fig:shuffle_frequencies}
\end{figure}

\begin{figure*}[t]
  \centering
  \includegraphics[width=\textwidth]{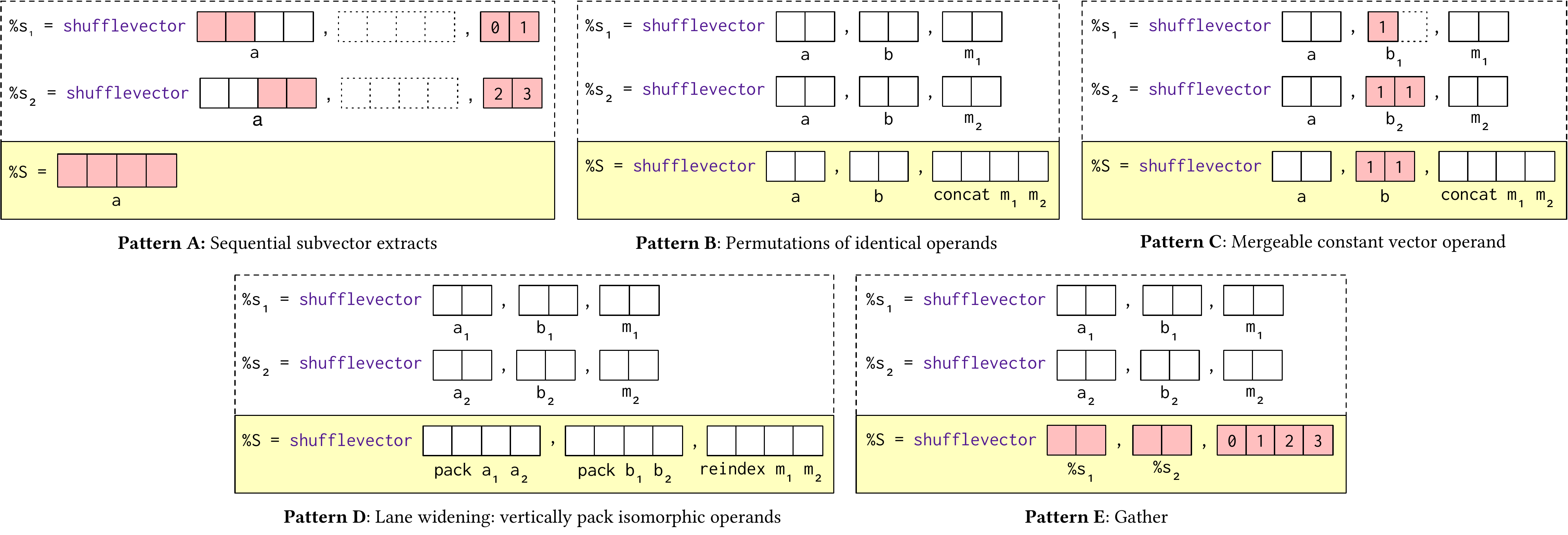}
    \caption{Shuffle pack transformation patterns. Revec avoids unnecessarily widening operands when applying patterns A, B and C. Pattern D, a lane widen, is applied when there is similarity between operands vertically in the pack, while narrow vectors are gathered only when an optimal packing decision cannot be made locally. Shaded operand elements are indexed by a shaded shuffle mask index.}
 \label{fig:shuffle_rules}
\end{figure*}

With shuffle merge rules (Figure \ref{fig:shuffle_rules}), \tool{} transforms this pack into a single wide shuffle $\texttt{S = shuffle A, B, M}$, where \texttt{A} and \texttt{B} are some combination of all \texttt{a\textsubscript{i}} and \texttt{b\textsubscript{i}}, and \texttt{M} is a statically determined constant vector. In Figure \ref{fig:shuffle_frequencies}, we show that four shuffle patterns match all packs encountered when revectorizing to AVX2, and 92.9\% of packs when revectorizing to AVX-512. While a gather or a generic lane widening strategy, Pattern D, could be applied to all packs, Patterns A, B, and C allow \tool{} to emit fewer or less costly shuffles.

\paragraph{Pattern A: Sequential Subvector Extraction} When $a_1 = a_2 = ... = a_p$, $|a_i| = n$, $|m_i| = \frac{n}{p}$,
and $\texttt{concat}~m_1~m_2~...~m_p = \{0, 1, ..., n - 1\}$, the narrow shuffles extract all adjacent $\frac{n}{p}$ length subvectors of source $a$ (operands $b_i$ are unused). \tool{} emits $\texttt{S} = a_1$ with no shuffling and erases the pack. Sequential subvector extraction is used by programmers to interleave subvectors from different source registers. By erasing the extractions, \tool{} can interleave the full source vectors elsewhere in the graph.

\paragraph{Pattern B: Permutations of Identical Operands} If $a_1 = a_2 = ... = a_p$, and $b_1 = b_2 = ... = b_p$, then \tool{} only widens the mask by concatenation.
The source operands need not be packed as a shuffle mask can repeat indices.

\paragraph{Pattern C: Mergeable Constant Operands}
Shuffles often include a constant vector as a source operand. For instance, the low and high halves of a source vector $a$ can be separately interleaved with a constant vector. In LLVM, the unused halves of the constant vector are in practice set to be undefined. If constant operands are non-identical but can be elementwise merged as in Figure \ref{fig:shuffle_rules}(C), and an operand is shared between narrow shuffles, \tool{} merges the constant and concatenates the masks.

\paragraph{Pattern D: Generic Lane Widening}
Failing the above conditions, \tool{} attempts to form pack $\alpha = \{a_1, a_2, ..., a_p\}$ from left operands and pack $\beta = \{b_1, b_2, ..., b_p\}$ from right operands.
The following local analyses indicate that a lane widening, vertical packing strategy is least costly:
\begin{itemize}[leftmargin=0.5cm]
    \item The left or right operand pack contains all constant vectors. \tool{} transforms the constant pack into a higher width constant vector.
    \item The left operands are identical or the right operands are identical. \tool{} can emit an efficient splat or broadcast instruction for this pack.
    \item All narrow masks are equal. The pack is isomorphic.
\end{itemize}

\tool{} generates a reindexed shuffle mask \texttt{M} by adding constant displacements to elements of each narrow mask $m_i$. Operands are displaced proportionally to their position in the pack. The constant displacement for indexes of narrow shuffle $i$ that correspond to the left operand (i.e indexes $<n$) is
$\sum_{j=1}^{i-1} |a_j| = n * (i - 1)$. Similarly, the constant displacement for indexes that correspond to values of the right operand is
$\sum_{j \neq i} |a_j|+ \sum_{j=1}^{i-1} |b_j| = n(p - 1) + n(i-1)$.

\paragraph{Gathering General Shuffles}
For shuffles with arbitrary non-constant operands, $2^{p-1}$ packs can be formed. As local packing decisions lead to revectorization graphs of different profitabilities, \tool{} is unable to make a globally optimal decision without a search. In the general case, narrow shuffles are gathered as described in Section \ref{sec:gather}. Any gathered shuffle pack will be non-isomorphic, as isomorphic shuffles have the same mask and match patterns B, C or D. \tool{} gathered no shuffle packs when revectorizing SSE2+ benchmarks to AVX2, and 7.1\% when revectorizing to AVX-512.

\subsection{Gathering Non-mergeable Vectors}
\label{sec:gather}
\begin{figure}[b]
  \includegraphics[width=\linewidth]{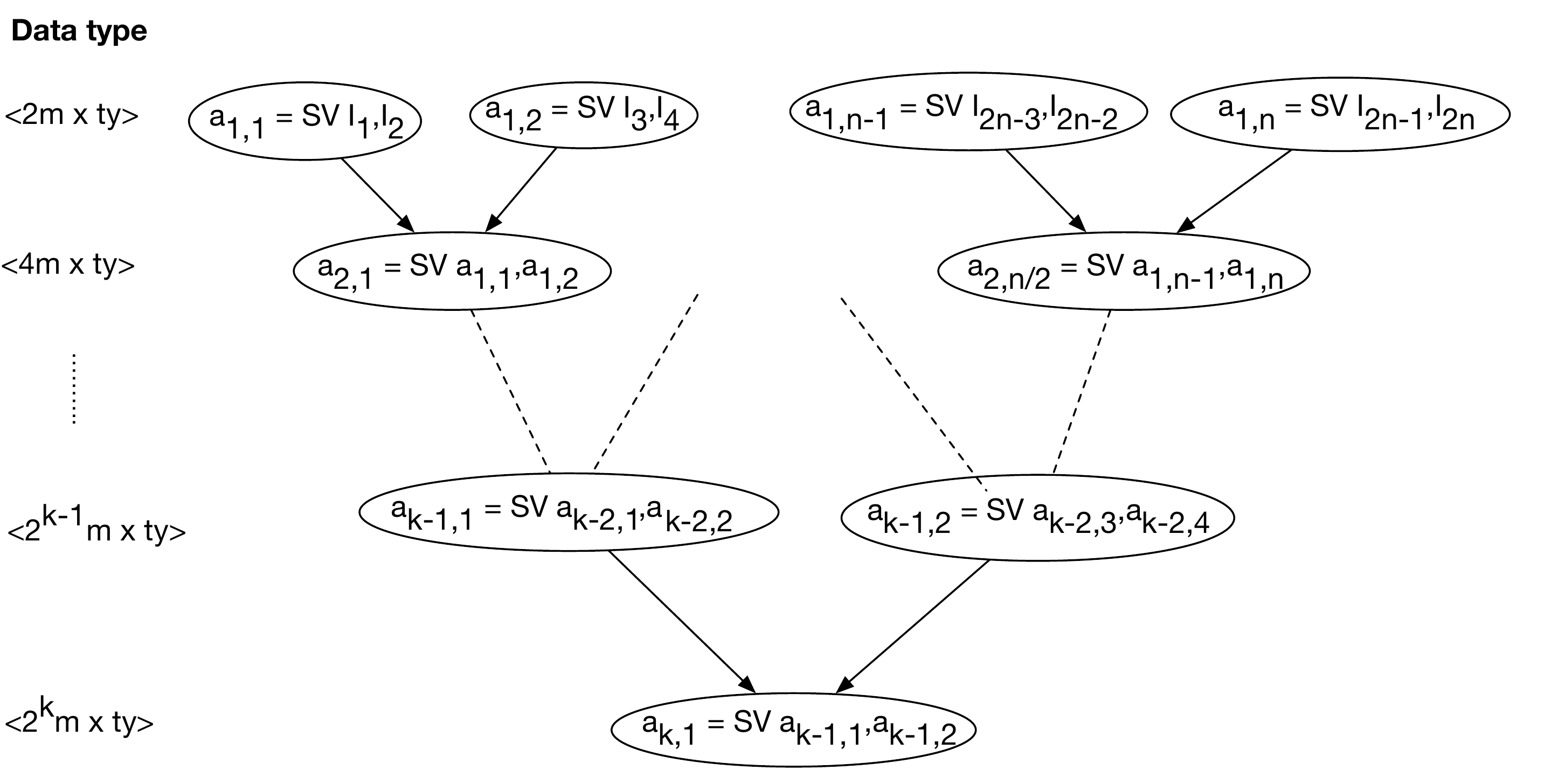}
    \caption{Tree of vector shuffles (SV) to gather packs which are not mergeable.}
  \label{fig:tree}
\end{figure}
\tool{} gathers packs at the leaf nodes of the graph if they are not mergeable. This is realized by using a tree of vector shuffles, where at each level it concatenates two vectors of the same width to form a vector of twice the data width.

Assume we want to merge a pack \{$I_1, I_2, ..., I_{2n}$\} of type \texttt{<m x ty>} instructions. The gathered wide vector of type \texttt{<(2n*m) x ty>} is formed using the tree of vector shuffles shown in Figure~\ref{fig:tree}. Vectors formed at level $j$ have $2^jm$ elements. The mask used for each shuffle at level $j$, $M_j = \texttt{<}0,1,2,....,2^jm\texttt{>}$. The height of the tree is $k = \log{2n} + 1$.

\section{Discovering Intrinsic Conversions}
\label{sec:enumeration}

In Section~\ref{sec:wide}, we noted that \tool{} queries an \texttt{IntrinsicMap} data structure to transform packs of opaque vector intrinsic calls.
\tool{} automatically generates intrinsic substitution candidates using explicit enumeration and pre-populates the \texttt{IntrinsicMap} structure used in Algorithm~\ref{alg:revec}.
When \texttt{IntrinsicMap(intrin,} $p$\texttt{) = wide\_intrin}, an isomorphic pack of $p$ calls to \texttt{intrin} has the same result as a single call to \texttt{wide\_intrin} with vertically packed operands. In our implementation of \tool{}, a canonical representation of each narrow intrinsic is mapped to several translations of higher width vector instructions at different vectorization factors $p$. We limit vectorization factors to powers of two.

We adopt a test case fuzzing approach for generating equivalences, with 18,000 randomly generated inputs and combinatorially generated corner-case inputs, inspired by test case generation in~\cite{ganesh2009taint}. Test cases are byte strings of a specified length that initialize operands of a specified vector data type.
To augment randomly generated test cases, we create corner-case byte string sequences such as $0000...$, $0101...$, $1010...$, and $1111...$. This helped \tool{} to significantly reduce erroneous equivalences, and is consistent with the methodology in~\cite{ganesh2009taint}.

For each vector intrinsic available on the target and each vectorization factor $1 \leq p \leq 8$, we generate and compile a testbed in LLVM IR that invokes the intrinsic $p$ times. Input test cases are passed as arguments of these intrinsics, and outputs are saved after execution. Testbeds are evaluated on a server with AVX-512 support (Section~\ref{ssec:eval_server}). If a set of testbeds produce the same output when given the same input for all test cases, each pair of testbeds in the set corresponds to a conversion. After filtering redundant conversions, the remaining conversions are added to \texttt{InstrinsicMap}.


Automated enumeration discovered 53 SSE-series to AVX1/2 intrinsic conversions, 33 AVX1/2 to AVX-512 conversions, and 19 SSE-series to AVX-512 conversions.
For instance, the SSE4.1 intrinsic \texttt{\_mm\_packus\_epi32} found in pack \textcircled{\small D} of Figure~\ref{fig:example} has a 2-to-1 conversion to \texttt{\_mm256\_packus\_epi32} and a 4-to-1 conversion to \texttt{\_mm512\_packus\_epi32}. Additionally, enumeration discovered several conversions that upgrade operand widths within the same instruction set.



\section{Implementation}
\label{sec:implementation}

We implemented \tool{} as a LLVM IR-level compiler pass extending LLVM v7.0.0 (pre-release, commit hash 71650da2). In addition, we add a loop analysis pass to determine inner loop unroll counts prior to revectorization. We execute the passes through Clang (commit hash 3ed5b23) as a frontend to compile C++ benchmarks. \tool{} transforms each source function immediately after the built-in SLP scalar autovectorization pass.

\section{Evaluation}
\label{sec:evaluation}

We evaluate the performance impact of \tool{} on real world performance critical kernels from video compression, integer packing, image processing, and stencil computation domains which contain heavily hand-vectorized routines.
In Section~\ref{ssec:eval_simdscan}, we examine \tool{}'s impact on integer unpacking routines \cite{willhalm2009simd} in the FastPFor integer compression library. In Section~\ref{ssec:eval_x265}, we evaluate \tool{}'s imapact on important kernels from the x265 video encoding library \cite{x265}. Finally, in Section~\ref{ssec:eval_simd}, we evaluate \tool{}'s impact on popular image processing and stencil kernels from the Simd image processing library~\cite{simd}.

\subsection{Experimental Setup}
\label{ssec:eval_server}
Benchmark kernels are run on a Google Cloud platform server with 4 vCPUs and 8GB of RAM. The benchmark server had Intel Xeon Skylake cores running at 2.00 GHz, and supported Intel's AVX-512 instructions (with the the F, CD, VL, DQ and BW instruction sets).
We target AVX2 by disabling AVX-512 features with the compilation flags { \small \texttt{-O3 -march=native -mno-avx512f -mno-avx512pf -mno-avx512er -mno-avx512cd}}.
We simply compile with { \small \texttt{-O3 -march=\allowbreak native}} to target AVX-512.
Test code measures absolute runtimes, and speedups are computed with respect to the runtime under compilation by stock Clang v7.0.0 (commit 3ed5b23) with the same command line flags, with no preprocessing or revectorization.

\subsection{Benchmark: \textsc{SIMD-Scan} Integer Unpacking}
\label{ssec:eval_simdscan}
\begin{figure}[t]
    \includegraphics[width=\linewidth]{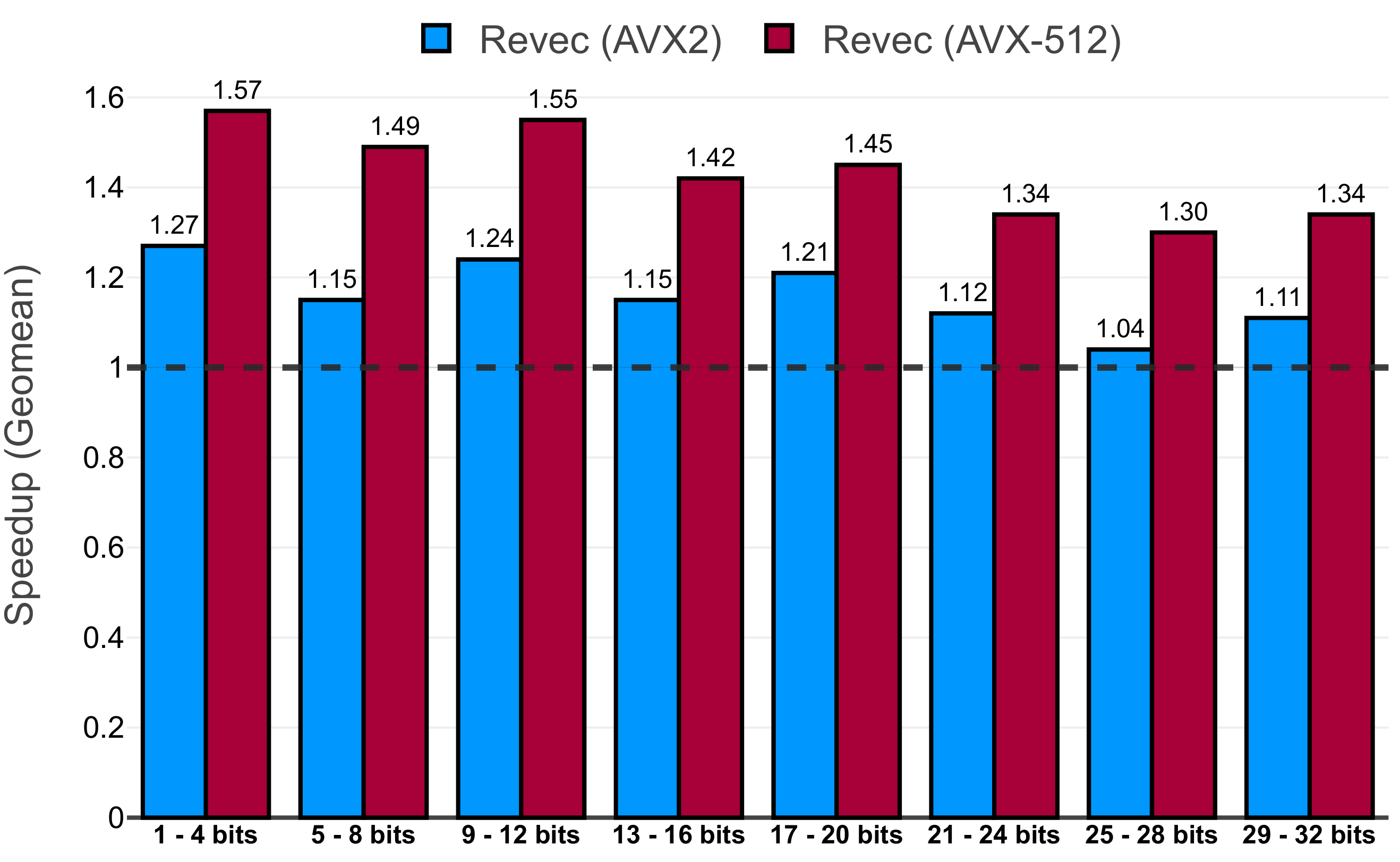}
    \caption{\tool{} speedups on 32 SSE4 horizontal bit unpacking kernels from \textsc{SIMD-Scan} and \textsc{FastPFor} when retargetted to AVX2 and AVX-512.}
    \label{fig:simdscan-speedup}
\end{figure}
\textsc{SIMD-Scan}~\cite{willhalm2009simd, willhalm2013vectorizing} proposes an efficient horizontal codeword-to-integer unpacking algorithm accelerated by 128-bit SIMD processors. Integer array packing and unpacking are widely used in search engine and relational database software to store data as codewords, so extensive effort has been spent on optimizing these routines \cite{lemire2015billions}. Decoding packed codewords into integers can be a bottleneck in lookup procedures, such as database column scans. The FastPFOR integer compression library \cite{lemire2015billions} implements 32 separate SSE4 unpacking kernels from \textsc{SIMD-Scan} for different integer bitwidths. Specialized kernels are used for each integer bitwidth as codewords packed to different bitwidths span SIMD register lanes differently.

In Figure \ref{fig:simdscan-speedup}, we report the speedups that \tool{} achieves on these kernels when retargeting to the AVX2 and AVX-512 instruction sets. While the authors of \cite{lemire2015billions} implement a separate function for each bitwidth, we aggregate speedups into groups of 4 for clarity.

Across the 32 SSE4 kernels, \tool{} nets a 1.160$\times$ geometric mean speedup over Clang when targeting AVX2. On the same experimental server, with AVX-512 instruction sets enabled, \tool{} nets a 1.430$\times$ geometric mean speedup.

Without \tool{}, Clang is unable to leverage newly available instruction sets given hand-vectorized code written to target the SSE4 instruction set. In fact, on this benchmark, we observe a slight slowdown from enabling AVX-512 instructions with stock Clang. However, \tool{} demonstrates the potential for automatic performance scaling of real-world code.

In fact, in later work, the \textsc{SIMD-Scan} authors propose a further optimized, AVX2-accelerated database column scan tool, \textsc{AVX2-Scan}, with horizontal unpacking being the key routine \cite{willhalm2013vectorizing}. The authors report that \textsc{AVX2-Scan} is "around 30\% faster than \textsc{SIMD-Scan}". \tool{} achieves considerable proportion of this proposed speedup automatically without manual intervention.

\subsection{Benchmark: x265 Video Encoding Library}
\label{ssec:eval_x265}
\begin{figure}[t]
    \includegraphics[width=0.88\linewidth]{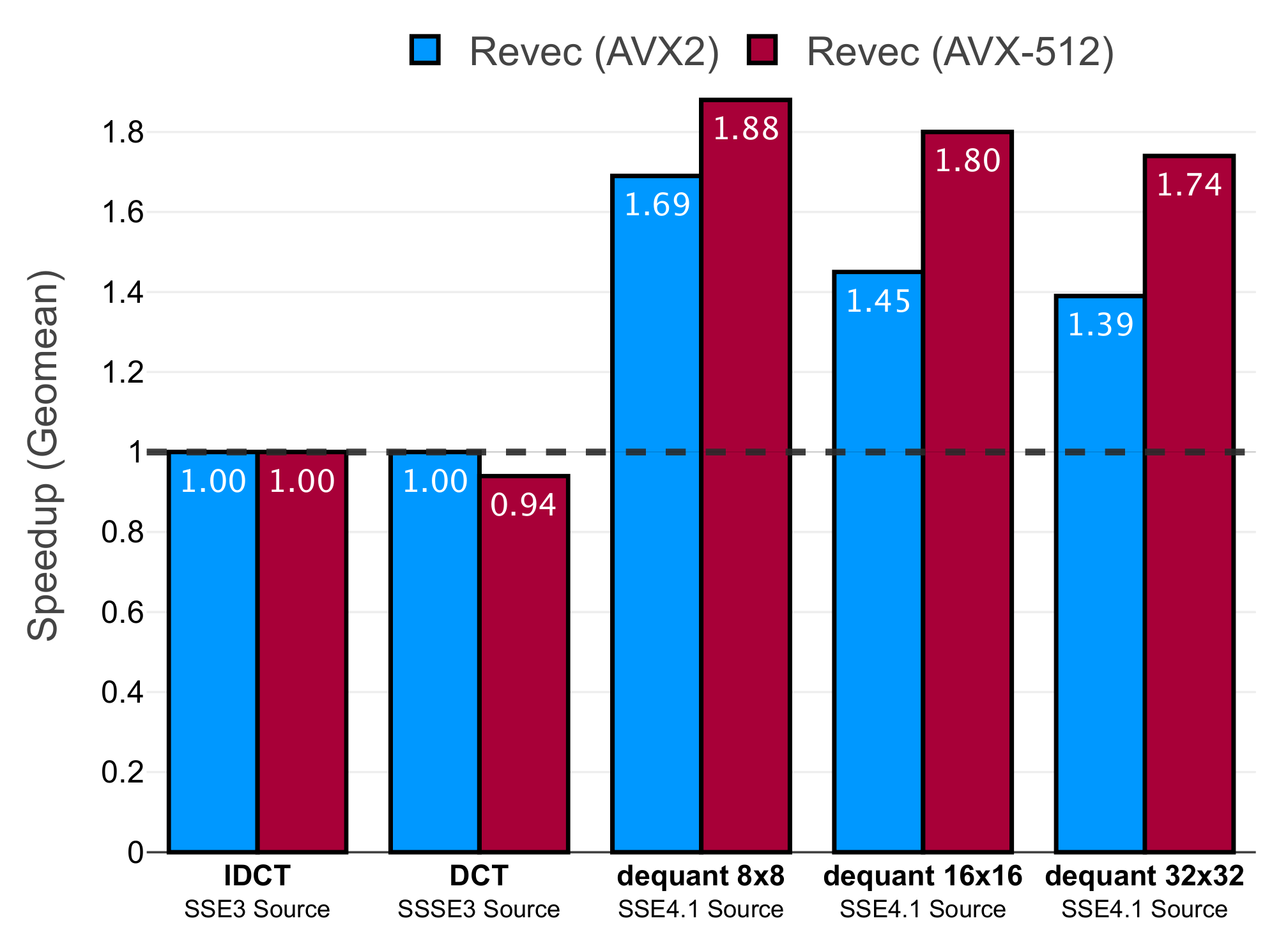}
    \caption{\tool{} speedups on hand-written SSE-series kernels used in the x265 video encoding library.}
    \label{fig:x265-speedup}
\end{figure}
x265 ~\cite{x265} is the most widely used HEVC/H.264 video encoding library. The library includes intrinsic-rich C++ routines for computing an inverse discrete cosine transform (IDCT) and discrete cosine transform (DCT) over small image blocks, and for image dequantization with scaling, accelerated with SSE-series intrinsics.

In Figure \ref{fig:x265-speedup}, we note \tool{}'s speedups on these routines. Overall, \tool{} achieves geometric mean speedups of 1.145$\times$ and 1.195$\times$ for AVX2 and AVX-512 targets, respectively.\footnote{Geometric mean taken over 3 numbers: DCT speedup, IDCT speedup, and aggregated dequantized scaling (dequant) speedup.}

 However, we observe a slight slowdown on the x265 DCT implementation for AVX-512 targets. \tool{} encounters packs of 4 SSSE3 \texttt{phadd} intrinsics. No 4-to-1 instrinsic equivalences are known for this instruction, as \texttt{phadd} is absent from AVX-512. While \tool{} currently gathers this pack with 3 shuffles (Section \ref{sec:gather}) such that an SSE-AVX switching penalty is incurred, \texttt{IntrinsicMap} contains a 2-to-1 equivalence to the AVX2 \texttt{phadd} intrinsic. The pack of 4 SSSE3 \texttt{phadd} intrinsics could be split and partially revectorized to use 2 AVX2 \texttt{phadd}s, gathered with a single shuffle.

 Further, the AVX2-retargetted DCT would benefit from horizontal packing of operands of an SSSE3 $\texttt{phadd}$ pack followed by data reorganization, as opposed to lane widening (vertical packing). Lane widening packs operands generated by different intrinsics that must be gathered and terminates the tree; horziontal packing would prevent this. \tool{}'s greedy algorithm misses this opportunity, however a search across packing strategies can be used uncover such uncommon opportunities.

\subsection{Benchmark: \textsc{Simd} Library}
\label{ssec:eval_simd}

\textsc{Simd}~\cite{simd} is a library of heavily hand-vectorized image processing and generic stencil kernels, with specialized implementations targetting different vector instruction sets. We group 216 SSE2+ kernels (168 SSE2, 47 SSE3/SSSE3, and 1 SSE4.1 kernels) based on their functionality in Table~\ref{tab:eval_simdfuncs}.

Each SSE-series kernel we extract and benchmark has an analogous AVX2 implementation. Hence, we can compare speedups of automatic revectorization with speedups that human experts have achieved. Other image processing libraries such as OpenCV provide a smaller variety of SIMD kernels. Further, the Intel IPP-ICV vector computer vision package does not provide source code.

\begin{table}[b]
    \caption{Classes of image processing and stencil kernels in the Simd library}
    \label{tab:eval_simdfuncs}
    \resizebox{\columnwidth}{!}{%
    \begin{tabular}{@{}lll@{}}
        \toprule
        \multicolumn{3}{c}{Simd library benchmark functions} \\ \midrule
        \textbf{Conversions} & \hspace{3mm}\texttt{reducegray5x5} & \textbf{Statistic} \\
        \hspace{3mm}\texttt{bgratogray} & \hspace{3mm}\texttt{shiftbilinear} & \hspace{3mm}\texttt{conditional} \\
        \hspace{3mm}\texttt{bgratoyuv} & \hspace{3mm}\texttt{stretchgray2x2} & \hspace{3mm}\texttt{statistic} \\
        \hspace{3mm}\texttt{bgrtobgra} &  &  \\
        \hspace{3mm}\texttt{bgrtoyuv} & \textbf{Filters} & \textbf{Correlation} \\
        \hspace{3mm}\texttt{deinterleave} & \hspace{3mm}\texttt{absgradientsaturatedsum} & \hspace{3mm}\texttt{absdifferencesum} \\
        \hspace{3mm}\texttt{graytobgra} & \hspace{3mm}\texttt{gaussianblur3x3} & \hspace{3mm}\texttt{squareddifferencesum} \\
        \hspace{3mm}\texttt{graytobgr} & \hspace{3mm}\texttt{laplace} &  \\
        \hspace{3mm}\texttt{int16togray} & \hspace{3mm}\texttt{lbp} & \textbf{Misc} \\
        \hspace{3mm}\texttt{interleave} & \hspace{3mm}\texttt{meanfilter3x3} & \hspace{3mm}\texttt{addfeaturedifference} \\
        \hspace{3mm}\texttt{yuvtobgra} & \hspace{3mm}\texttt{sobel} & \hspace{3mm}\texttt{binarization} \\
        \hspace{3mm}\texttt{yuvtobgr} &  & \hspace{3mm}\texttt{fill} \\
        \hspace{3mm}\texttt{yuvtohue} & \textbf{Motion detection} & \hspace{3mm}\texttt{histogram} \\
         & \hspace{3mm}\texttt{background} & \hspace{3mm}\texttt{hog} \\
        \textbf{Resizing} & \hspace{3mm}\texttt{edgebackground} & \hspace{3mm}\texttt{neural} \\
        \hspace{3mm}\texttt{reducegray2x2} & \hspace{3mm}\texttt{interference} & \hspace{3mm}\texttt{operation} \\
        \hspace{3mm}\texttt{reducegray3x3} & \hspace{3mm}\texttt{segmentation} & \hspace{3mm}\texttt{reorder} \\
        \hspace{3mm}\texttt{reducegray4x4} & \hspace{3mm}\texttt{texture} &  \\ \bottomrule
    \end{tabular}
    }
\end{table}

\subsubsection{Impact of \tool{} on Runtime Performance}

\begin{figure*}[t]
    \includegraphics[width=\textwidth]{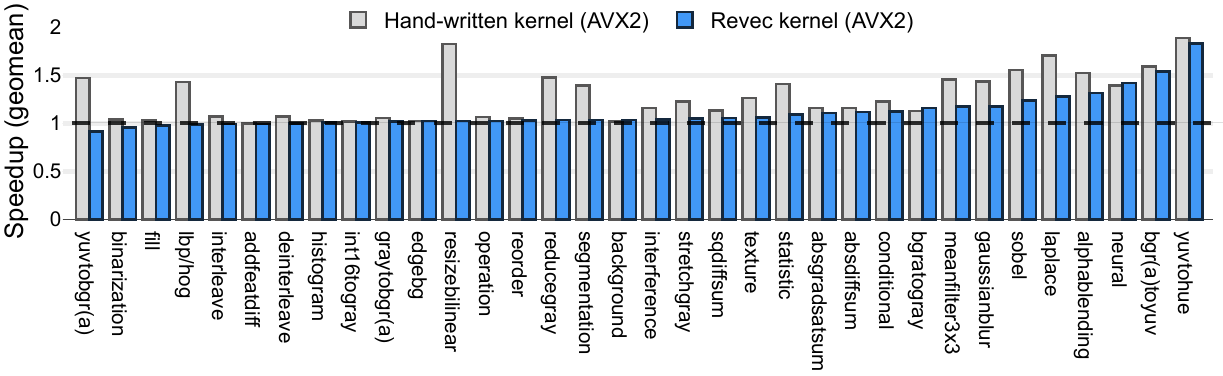}
    \includegraphics[width=\textwidth]{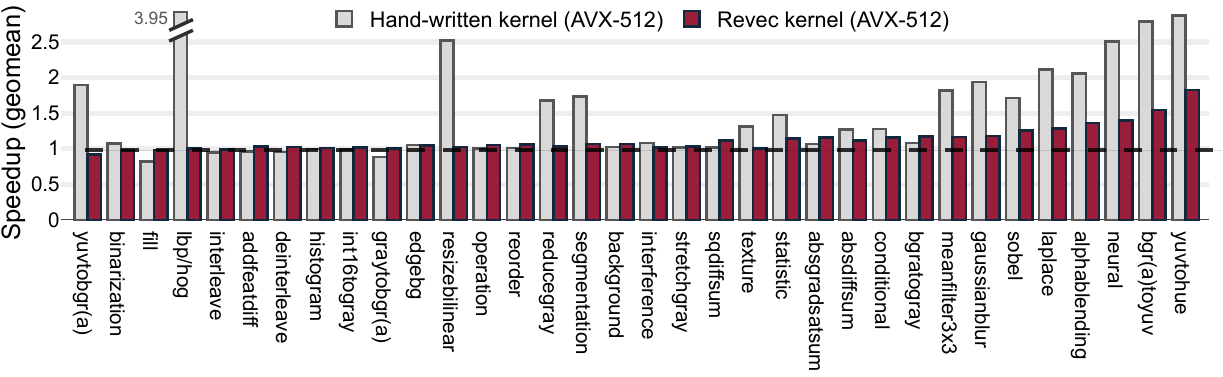}
    \caption{Revectorized and hand-written AVX2 kernel speedups (top) and AVX-512 speedups (bottom) over hand-written SSE2+ kernels for various \textsc{Simd} library benchmarks}
    \label{fig:speedup}
\end{figure*}

We report the speedup \tool{} has over SSE2+ implementations of \textsc{Simd} library kernels when retargeted to produce AVX2 or AVX-512 code in Figure~\ref{fig:speedup}. Note that for presentation purposes, we clustered the kernels into the classes shown in Table~\ref{tab:eval_simdfuncs}. In Figure~\ref{fig:speedup}, the speedup for each cluster is the geometric mean speedup of all kernels within a cluster. We benchmark with test code written by the \textsc{Simd} authors that executes each kernel with large inputs (i.e. 1920 x 1080 pixels and other similar sizes).

Across 216 extracted, templatized variants of SSE2+ \textsc{Simd} library kernels from 2016, \tool{} achieves an overall geometric mean speedup of \textbf{1.1023$\times$} when retargeting to AVX2, and \textbf{1.1163$\times$} when retargeting to AVX-512. In comparison, handwritten AVX2 implementations from the same commit have a 1.2673$\times$ speedup over their SSE2+ counterparts. After we extracted kernels, the \textsc{Simd} library authors added handwritten AVX-512 implementations. From a December 2018 commit of the \textsc{Simd} library, handwritten AVX-512 kernels had a 1.3916$\times$ speedup over SSE2+ counterparts.

We exclude 14 kernels that achieve less than a 1.01 speedup when both hand-rewritten and revectorized to AVX2 and AVX-512 --- these are memory-bound functions with limited opportunity for revectorization. Nonetheless, elided kernels also compile and pass tests with neutral performance.

In addition, on 6 SSSE3/SSE2 color conversion tests in the \textsc{Simd} library that have no AVX2 or AVX-512 implementations to compare against, \tool{} yields a 1.0618$\times$ and 1.0632$\times$ geometric mean speedup over stock Clang when retargeting to AVX2 and AVX-512 respectively.

\subsubsection{Comparison with Hand-Vectorized Kernels}

While hand-written AVX2 and AVX-512 \textsc{Simd} kernels overall perform better than revectorized kernels, these required considerable manual effort and expertise to write.
In particular, hand-vectorization achieves >3$\times$ speedup on one kernel variant in the \texttt{resizebi\-linear} cluster. This large speedup is due to the usage of \texttt{\_mm256\allowbreak\_maddubs\allowbreak\_epi16} in the AVX2 implementation, which executes vertical multiplies and horizontal additions. The SSE2 version emulates this using multiple intrinsics involving vector shuffles. Such complex semantic transformations are not currently possible under \tool{}. Similarly, \tool{} can only generate the horizontal vector operation \texttt{\_mm256\_hadd\_epi32} given SSE3-onward source; SSE2 sources emulate the function. Horizontal AVX2 operations are used in other hand-written AVX2 kernels such as \texttt{reducegray4x4}.

The AVX-512 implementation of the \texttt{hog} kernel uses fused multiply and add intrinsics while the SSE2 source, written to target processors without the FMA instruction set, separately multiplies and adds. The AVX-512 \texttt{hog} and \texttt{lbp} kernels also use masked selects or sets like \texttt{\_mm512\_mask\_blend\_\allowbreak epi32}, whereas \tool{} widens casts, logical operations, or compares that emulate masked operations.

Many of these performance gaps could be narrowed by applying peephole-style conversion rules independently of the \tool{} transformation pass.

\tool{} still automatically delivers an overall speedup on \textsc{Simd}. Notably,
the \texttt{NeuralConvert} kernel enjoys a 1.757$\times$ speedup via revectorization to AVX2, but only 1.075$\times$ via hand vectorization. In the AVX2 kernel, \textsc{Simd} programmers reduced the number of unpacks/shuffles executed by using an AVX2-only zero extension intrinsic, \texttt{\_mm256\_cvtepu8\_epi32}, at the cost of using narrower loads (\texttt{load <8 x i8>}). The SSE2 implementation uses wider loads (\texttt{load <16 x i8>}), with more data rearrangement.
By revectorizing, \tool{} doubles the width of the high-latency loads to \texttt{<32 x i8>}, and uses our shuffle logic to merge unpacks. Revectorized AVX2 utilizes memory bandwidth better, even though it executes more instructions, yielding a higher speedup.

\section{Related Work}

Revectorization is inspired by compiler auto-vectorization. Also, previous works perform limited revectorization, primarily outside of compilers.
\paragraph{Compiler Auto-vectorization} Compilers employ two main auto-vectorization techniques, namely loop vectorization and Superword Level Parallelism (SLP) based vectorization. Loop vectorization has been implemented in compilers since the era of vector machines~\cite{FortranLoop} and subsequently many vectorization schemes have been proposed which use loop dependency analysis~\cite{vlp01}. Other loop vectorization techniques explore vectorization under alignment constraints~\cite{alignment}, outer loop transformations~\cite{outerLoop}, handling data interleavings in loops~\cite{interleave} and exploiting mixed SIMD parallelism~\cite{dataReorg, reorgLarsen}.

Larsen~\cite{LarsenSLP} introduced SLP, which can capture vectorization opportunities that exist beyond loops at a much lower granularity. \tool{} has its roots in SLP vectorization which packs isomorphic independent statements starting from scalar loads and stores to form vectorized counterparts. In particular, \tool{}'s transformation algorithm is inspired by a variant of SLP proposed in~\cite{Rosen2007}, but targets only vector instructions.

Domain specific vectorization techniques~\cite{domain-franz,fft-franz} have been proposed in systems like SPIRAL~\cite{spiraln} where generic auto-vectorization fails. They use platform-specific vector intrinsics in their implementations and \tool{} can be used to rejuvenate their performance on newer instructions sets.

\paragraph{Dynamic Rewriting} Dynamic rewriting of SIMD instructions has been proposed in~\cite{dyn-simd,hqemu} to find SIMD mappings between host and guest architectures in dynamic binary translation. The Dynamic Binary Translation system proposed in~\cite{dyn-revec} details a technique to widen SIMD instructions during this mapping. It only targets loops and relies on recovery of LLVM IR from the binary, which is imprecise and can lead to spurious dependencies. Compared to~\cite{dyn-revec},~\tool{} is implemented as a compiler level transformation pass and inherently has access to loop structures without the need to recover them from the binary, making \tool{} more precise. Further,~\tool{} applies to loop-free segments of code, making it more general than \cite{dyn-revec} -- the Simd \texttt{NeuralConvert} benchmark that \tool{} greatly accelerates depends on this capability.

\paragraph{Static Rewriting} Manilov~\cite{free-rider} proposes a source-to-source translation system that maps intrinsic code written in one platform to another. It uses intrinsic descriptions written separately in header files for two platforms to match dataflow graphs between intrinsics to do the translation. It can be used for widening vector intrinsics as well similar to \tool{}. However, it requires considerable manual engineering: semantically correct descriptions should be written for each vector intrinsic in plain C code. This is tedious and error prone. In comparison, \tool{} automatically finds these equivalences using enumeration. Further, \tool{} is implemented as a compiler pass and transformation happens transparently without the need for library modification: \tool{} can directly benefit end-users of software. Pokam~\cite{swarp} proposes a system which analyzes C-like code and finds common idioms which are vectorizable and translates those to SIMD intrinsics.~\tool{} analyzes code already written using vector intrinsics.

Libraries like~\cite{generic-simd,vc} provide abstract ways of programming SIMD operations across platforms. These systems are suitable for writing new applications, but do not deal with the portability of high performance applications already written using platform-specific vector intrinsics.

\section{Conclusion}
\label{sec:conclusion}

Programmers write code using low-level platform-specific vector intrinsics to exploit data level parallelism as much as possible. However, hand vectorization results in non-portable code, and hence programmers resort to manually rewriting code using intrinsics of newer vector instruction sets to leverage the availability of higher vector width instructions. In this paper, we introduced \tool{}, a compiler technique which automatically retargets hand vectorized code to use higher vector width instructions whenever available. We showed \tool{} can be used to transparently rejuvenate performance of stale hand vectorized code, thereby achieving performance portability.

\balance

\begin{acks}

We would like to thank Vladimir Kiriansky and all reviewers for insightful comments and suggestions.
This work was supported by the U.S. Department of Energy, Office of Science, Office of Advanced Scientific Computing Research under Award Numbers DE-SC0008923 and DE-SC0018121; the National Science Foundation under Grant No. CCF-1533753; and DARPA under Award Number HR0011-18-3-0007. Any opinions, findings, and conclusions or recommendations expressed in this material are those of the authors and do not necessarily reflect the views of the funding agencies.

\end{acks}

\bibliography{citations.bib}


\begin{thebibliography}{31}


\ifx \showCODEN    \undefined \def \showCODEN     #1{\unskip}     \fi
\ifx \showDOI      \undefined \def \showDOI       #1{#1}\fi
\ifx \showISBNx    \undefined \def \showISBNx     #1{\unskip}     \fi
\ifx \showISBNxiii \undefined \def \showISBNxiii  #1{\unskip}     \fi
\ifx \showISSN     \undefined \def \showISSN      #1{\unskip}     \fi
\ifx \showLCCN     \undefined \def \showLCCN      #1{\unskip}     \fi
\ifx \shownote     \undefined \def \shownote      #1{#1}          \fi
\ifx \showarticletitle \undefined \def \showarticletitle #1{#1}   \fi
\ifx \showURL      \undefined \def \showURL       {\relax}        \fi
\providecommand\bibfield[2]{#2}
\providecommand\bibinfo[2]{#2}
\providecommand\natexlab[1]{#1}
\providecommand\showeprint[2][]{arXiv:#2}

\bibitem[\protect\citeauthoryear{Allen and Kennedy}{Allen and Kennedy}{1987}]%
        {FortranLoop}
\bibfield{author}{\bibinfo{person}{Randy Allen} {and} \bibinfo{person}{Ken
  Kennedy}.} \bibinfo{year}{1987}\natexlab{}.
\newblock \showarticletitle{Automatic Translation of FORTRAN Programs to Vector
  Form}.
\newblock \bibinfo{journal}{\emph{ACM Trans. Program. Lang. Syst.}}
  \bibinfo{volume}{9}, \bibinfo{number}{4} (\bibinfo{date}{Oct.}
  \bibinfo{year}{1987}), \bibinfo{pages}{491--542}.
\newblock
\showISSN{0164-0925}
\urldef\tempurl%
\url{https://doi.org/10.1145/29873.29875}
\showDOI{\tempurl}


\bibitem[\protect\citeauthoryear{ARM}{ARM}{2013}]%
        {neon}
\bibfield{author}{\bibinfo{person}{ARM}.} \bibinfo{year}{2013}\natexlab{}.
\newblock \bibinfo{title}{ARM Programmer Guide}.
\newblock
\newblock
\urldef\tempurl%
\url{http://infocenter.arm.com/help/index.jsp?topic=/com.arm.doc.den0018a/index.html}
\showURL{%
\tempurl}


\bibitem[\protect\citeauthoryear{Eichenberger, Wu, and O'Brien}{Eichenberger
  et~al\mbox{.}}{2004}]%
        {alignment}
\bibfield{author}{\bibinfo{person}{Alexandre~E. Eichenberger},
  \bibinfo{person}{Peng Wu}, {and} \bibinfo{person}{Kevin O'Brien}.}
  \bibinfo{year}{2004}\natexlab{}.
\newblock \showarticletitle{Vectorization for SIMD Architectures with Alignment
  Constraints}. In \bibinfo{booktitle}{\emph{Proceedings of the ACM SIGPLAN
  2004 Conference on Programming Language Design and Implementation}}
  \emph{(\bibinfo{series}{PLDI '04})}. \bibinfo{publisher}{ACM},
  \bibinfo{address}{New York, NY, USA}, \bibinfo{pages}{82--93}.
\newblock
\showISBNx{1-58113-807-5}
\urldef\tempurl%
\url{https://doi.org/10.1145/996841.996853}
\showDOI{\tempurl}


\bibitem[\protect\citeauthoryear{Franchetti, Kral, Lorenz, and
  Ueberhuber}{Franchetti et~al\mbox{.}}{2005}]%
        {domain-franz}
\bibfield{author}{\bibinfo{person}{Franz Franchetti}, \bibinfo{person}{Stefan
  Kral}, \bibinfo{person}{Juergen Lorenz}, {and} \bibinfo{person}{Christoph~W
  Ueberhuber}.} \bibinfo{year}{2005}\natexlab{}.
\newblock \showarticletitle{Efficient utilization of SIMD extensions}.
\newblock \bibinfo{journal}{\emph{Proc. IEEE}} \bibinfo{volume}{93},
  \bibinfo{number}{2} (\bibinfo{year}{2005}), \bibinfo{pages}{409--425}.
\newblock


\bibitem[\protect\citeauthoryear{Fu, Hong, Wu, Liu, and Hsu}{Fu
  et~al\mbox{.}}{2015}]%
        {hqemu}
\bibfield{author}{\bibinfo{person}{S. Fu}, \bibinfo{person}{D. Hong},
  \bibinfo{person}{J. Wu}, \bibinfo{person}{P. Liu}, {and} \bibinfo{person}{W.
  Hsu}.} \bibinfo{year}{2015}\natexlab{}.
\newblock \showarticletitle{SIMD Code Translation in an Enhanced HQEMU}. In
  \bibinfo{booktitle}{\emph{2015 IEEE 21st International Conference on Parallel
  and Distributed Systems (ICPADS)}}. \bibinfo{pages}{507--514}.
\newblock
\showISSN{1521-9097}
\urldef\tempurl%
\url{https://doi.org/10.1109/ICPADS.2015.70}
\showDOI{\tempurl}


\bibitem[\protect\citeauthoryear{Ganesh, Leek, and Rinard}{Ganesh
  et~al\mbox{.}}{2009}]%
        {ganesh2009taint}
\bibfield{author}{\bibinfo{person}{Vijay Ganesh}, \bibinfo{person}{Tim Leek},
  {and} \bibinfo{person}{Martin Rinard}.} \bibinfo{year}{2009}\natexlab{}.
\newblock \showarticletitle{Taint-based directed whitebox fuzzing}. In
  \bibinfo{booktitle}{\emph{Proceedings of the 31st International Conference on
  Software Engineering}}. IEEE Computer Society, \bibinfo{pages}{474--484}.
\newblock


\bibitem[\protect\citeauthoryear{Hong, Liu, Fu, Wu, and Hsu}{Hong
  et~al\mbox{.}}{2018}]%
        {dyn-revec}
\bibfield{author}{\bibinfo{person}{Ding-Yong Hong}, \bibinfo{person}{Yu-Ping
  Liu}, \bibinfo{person}{Sheng-Yu Fu}, \bibinfo{person}{Jan-Jan Wu}, {and}
  \bibinfo{person}{Wei-Chung Hsu}.} \bibinfo{year}{2018}\natexlab{}.
\newblock \showarticletitle{Improving SIMD Parallelism via Dynamic Binary
  Translation}.
\newblock \bibinfo{journal}{\emph{ACM Trans. Embed. Comput. Syst.}}
  \bibinfo{volume}{17}, \bibinfo{number}{3}, Article \bibinfo{articleno}{61}
  (\bibinfo{date}{Feb.} \bibinfo{year}{2018}), \bibinfo{numpages}{27}~pages.
\newblock
\showISSN{1539-9087}
\urldef\tempurl%
\url{https://doi.org/10.1145/3173456}
\showDOI{\tempurl}


\bibitem[\protect\citeauthoryear{IBM}{IBM}{2006}]%
        {vmx}
\bibfield{author}{\bibinfo{person}{IBM}.} \bibinfo{year}{2006}\natexlab{}.
\newblock \showarticletitle{PowerPC microprocessor family: Vector/SIMD
  multimedia extension technology programming environments manual}.
\newblock \bibinfo{journal}{\emph{IBM Systems and Technology Group}}
  (\bibinfo{year}{2006}).
\newblock


\bibitem[\protect\citeauthoryear{Ihar, Mikhail, Andrey, Fedorov, and
  Matsaberydze}{Ihar et~al\mbox{.}}{2016}]%
        {simd}
\bibfield{author}{\bibinfo{person}{Yermalayeu Ihar}, \bibinfo{person}{Antonenka
  Mikhail}, \bibinfo{person}{Radchenko Andrey}, \bibinfo{person}{Dmitry
  Fedorov}, {and} \bibinfo{person}{Kirill Matsaberydze}.}
  \bibinfo{year}{2016}\natexlab{}.
\newblock \bibinfo{title}{Simd Library for Image Processing}.
\newblock
\newblock
\urldef\tempurl%
\url{http://ermig1979.github.io/Simd/index.html}
\showURL{%
\tempurl}


\bibitem[\protect\citeauthoryear{Inc.}{Inc.}{2018}]%
        {x265}
\bibfield{author}{\bibinfo{person}{MulticoreWare Inc.}}
  \bibinfo{year}{2018}\natexlab{}.
\newblock \bibinfo{title}{x265 HEVC Encoder / H.265 Video Codec}.
\newblock
\newblock
\urldef\tempurl%
\url{http://x265.org}
\showURL{%
\tempurl}


\bibitem[\protect\citeauthoryear{Kretz and Lindenstruth}{Kretz and
  Lindenstruth}{2012}]%
        {vc}
\bibfield{author}{\bibinfo{person}{Matthias Kretz} {and}
  \bibinfo{person}{Volker Lindenstruth}.} \bibinfo{year}{2012}\natexlab{}.
\newblock \showarticletitle{Vc: A C++ library for explicit vectorization}.
\newblock \bibinfo{journal}{\emph{Software: Practice and Experience}}
  \bibinfo{volume}{42}, \bibinfo{number}{11}, \bibinfo{pages}{1409--1430}.
\newblock


\bibitem[\protect\citeauthoryear{Larsen and Amarasinghe}{Larsen and
  Amarasinghe}{2000}]%
        {LarsenSLP}
\bibfield{author}{\bibinfo{person}{Samuel Larsen} {and} \bibinfo{person}{Saman
  Amarasinghe}.} \bibinfo{year}{2000}\natexlab{}.
\newblock \showarticletitle{Exploiting Superword Level Parallelism with
  Multimedia Instruction Sets}. In \bibinfo{booktitle}{\emph{Proceedings of the
  ACM SIGPLAN 2000 Conference on Programming Language Design and
  Implementation}} \emph{(\bibinfo{series}{PLDI '00})}.
  \bibinfo{publisher}{ACM}, \bibinfo{address}{New York, NY, USA},
  \bibinfo{pages}{145--156}.
\newblock
\showISBNx{1-58113-199-2}
\urldef\tempurl%
\url{https://doi.org/10.1145/349299.349320}
\showDOI{\tempurl}


\bibitem[\protect\citeauthoryear{Larsen, Witchel, and Amarasinghe}{Larsen
  et~al\mbox{.}}{2002}]%
        {reorgLarsen}
\bibfield{author}{\bibinfo{person}{Samuel Larsen}, \bibinfo{person}{Emmett
  Witchel}, {and} \bibinfo{person}{Saman~P. Amarasinghe}.}
  \bibinfo{year}{2002}\natexlab{}.
\newblock \showarticletitle{Increasing and Detecting Memory Address
  Congruence}. In \bibinfo{booktitle}{\emph{Proceedings of the 2002
  International Conference on Parallel Architectures and Compilation
  Techniques}} \emph{(\bibinfo{series}{PACT '02})}. \bibinfo{publisher}{IEEE
  Computer Society}, \bibinfo{address}{Washington, DC, USA},
  \bibinfo{pages}{18--29}.
\newblock
\showISBNx{0-7695-1620-3}
\urldef\tempurl%
\url{http://dl.acm.org/citation.cfm?id=645989.674329}
\showURL{%
\tempurl}


\bibitem[\protect\citeauthoryear{Lattner and Adve}{Lattner and Adve}{2004}]%
        {LLVM}
\bibfield{author}{\bibinfo{person}{Chris Lattner} {and} \bibinfo{person}{Vikram
  Adve}.} \bibinfo{year}{2004}\natexlab{}.
\newblock \showarticletitle{LLVM: A Compilation Framework for Lifelong Program
  Analysis \& Transformation}. In \bibinfo{booktitle}{\emph{Proceedings of the
  International Symposium on Code Generation and Optimization:
  Feedback-directed and Runtime Optimization}} \emph{(\bibinfo{series}{CGO
  '04})}. \bibinfo{publisher}{IEEE Computer Society},
  \bibinfo{address}{Washington, DC, USA}, \bibinfo{pages}{75--}.
\newblock
\showISBNx{0-7695-2102-9}
\urldef\tempurl%
\url{http://dl.acm.org/citation.cfm?id=977395.977673}
\showURL{%
\tempurl}


\bibitem[\protect\citeauthoryear{Lemire and Boytsov}{Lemire and
  Boytsov}{2015}]%
        {lemire2015billions}
\bibfield{author}{\bibinfo{person}{Daniel Lemire} {and} \bibinfo{person}{Leonid
  Boytsov}.} \bibinfo{year}{2015}\natexlab{}.
\newblock \showarticletitle{Decoding billions of integers per second through
  vectorization}.
\newblock \bibinfo{journal}{\emph{Journal of Software Practice and Experience}}
  (\bibinfo{year}{2015}).
\newblock


\bibitem[\protect\citeauthoryear{Li, Zhang, Xu, and Huang}{Li
  et~al\mbox{.}}{2006}]%
        {dyn-simd}
\bibfield{author}{\bibinfo{person}{Jianhui Li}, \bibinfo{person}{Qi Zhang},
  \bibinfo{person}{Shu Xu}, {and} \bibinfo{person}{Bo Huang}.}
  \bibinfo{year}{2006}\natexlab{}.
\newblock \showarticletitle{Optimizing dynamic binary translation for SIMD
  instructions}. In \bibinfo{booktitle}{\emph{International Symposium on Code
  Generation and Optimization (CGO'06)}}. \bibinfo{pages}{12 pp.--280}.
\newblock
\urldef\tempurl%
\url{https://doi.org/10.1109/CGO.2006.27}
\showDOI{\tempurl}


\bibitem[\protect\citeauthoryear{Maleki, Gao, Garzar, Wong, Padua,
  et~al\mbox{.}}{Maleki et~al\mbox{.}}{2011}]%
        {maleki2011evaluation}
\bibfield{author}{\bibinfo{person}{Saeed Maleki}, \bibinfo{person}{Yaoqing
  Gao}, \bibinfo{person}{Maria~J Garzar}, \bibinfo{person}{Tommy Wong},
  \bibinfo{person}{David~A Padua}, {et~al\mbox{.}}}
  \bibinfo{year}{2011}\natexlab{}.
\newblock \showarticletitle{An evaluation of vectorizing compilers}. In
  \bibinfo{booktitle}{\emph{Parallel Architectures and Compilation Techniques
  (PACT), 2011 International Conference on}}. IEEE, \bibinfo{pages}{372--382}.
\newblock


\bibitem[\protect\citeauthoryear{Manilov, Franke, Magrath, and Andrieu}{Manilov
  et~al\mbox{.}}{2015}]%
        {free-rider}
\bibfield{author}{\bibinfo{person}{Stanislav Manilov},
  \bibinfo{person}{Bj\"{o}rn Franke}, \bibinfo{person}{Anthony Magrath}, {and}
  \bibinfo{person}{Cedric Andrieu}.} \bibinfo{year}{2015}\natexlab{}.
\newblock \showarticletitle{Free Rider: A Tool for Retargeting
  Platform-Specific Intrinsic Functions}.
\newblock \bibinfo{journal}{\emph{SIGPLAN Not.}} \bibinfo{volume}{50},
  \bibinfo{number}{5}, Article \bibinfo{articleno}{5} (\bibinfo{date}{June}
  \bibinfo{year}{2015}), \bibinfo{numpages}{10}~pages.
\newblock
\showISSN{0362-1340}
\urldef\tempurl%
\url{https://doi.org/10.1145/2808704.2754962}
\showDOI{\tempurl}


\bibitem[\protect\citeauthoryear{McFarlin, Arbatov, Franchetti, and
  P\"{u}schel}{McFarlin et~al\mbox{.}}{2011}]%
        {fft-franz}
\bibfield{author}{\bibinfo{person}{Daniel~S. McFarlin},
  \bibinfo{person}{Volodymyr Arbatov}, \bibinfo{person}{Franz Franchetti},
  {and} \bibinfo{person}{Markus P\"{u}schel}.} \bibinfo{year}{2011}\natexlab{}.
\newblock \showarticletitle{Automatic SIMD Vectorization of Fast Fourier
  Transforms for the Larrabee and AVX Instruction Sets}. In
  \bibinfo{booktitle}{\emph{Proceedings of the International Conference on
  Supercomputing}} \emph{(\bibinfo{series}{ICS '11})}.
  \bibinfo{publisher}{ACM}, \bibinfo{address}{New York, NY, USA},
  \bibinfo{pages}{265--274}.
\newblock
\showISBNx{978-1-4503-0102-2}
\urldef\tempurl%
\url{https://doi.org/10.1145/1995896.1995938}
\showDOI{\tempurl}


\bibitem[\protect\citeauthoryear{Nuzman, Rosen, and Zaks}{Nuzman
  et~al\mbox{.}}{2006}]%
        {interleave}
\bibfield{author}{\bibinfo{person}{Dorit Nuzman}, \bibinfo{person}{Ira Rosen},
  {and} \bibinfo{person}{Ayal Zaks}.} \bibinfo{year}{2006}\natexlab{}.
\newblock \showarticletitle{Auto-vectorization of Interleaved Data for SIMD}.
  In \bibinfo{booktitle}{\emph{Proceedings of the 27th ACM SIGPLAN Conference
  on Programming Language Design and Implementation}}
  \emph{(\bibinfo{series}{PLDI '06})}. \bibinfo{publisher}{ACM},
  \bibinfo{address}{New York, NY, USA}, \bibinfo{pages}{132--143}.
\newblock
\showISBNx{1-59593-320-4}
\urldef\tempurl%
\url{https://doi.org/10.1145/1133981.1133997}
\showDOI{\tempurl}


\bibitem[\protect\citeauthoryear{Nuzman and Zaks}{Nuzman and Zaks}{2008}]%
        {outerLoop}
\bibfield{author}{\bibinfo{person}{Dorit Nuzman} {and} \bibinfo{person}{Ayal
  Zaks}.} \bibinfo{year}{2008}\natexlab{}.
\newblock \showarticletitle{Outer-loop Vectorization: Revisited for Short SIMD
  Architectures}. In \bibinfo{booktitle}{\emph{Proceedings of the 17th
  International Conference on Parallel Architectures and Compilation
  Techniques}} \emph{(\bibinfo{series}{PACT '08})}. \bibinfo{publisher}{ACM},
  \bibinfo{address}{New York, NY, USA}, \bibinfo{pages}{2--11}.
\newblock
\showISBNx{978-1-60558-282-5}
\urldef\tempurl%
\url{https://doi.org/10.1145/1454115.1454119}
\showDOI{\tempurl}


\bibitem[\protect\citeauthoryear{Oberman, Favor, and Weber}{Oberman
  et~al\mbox{.}}{1999}]%
        {amd3dnow}
\bibfield{author}{\bibinfo{person}{Stuart Oberman}, \bibinfo{person}{Greg
  Favor}, {and} \bibinfo{person}{Fred Weber}.} \bibinfo{year}{1999}\natexlab{}.
\newblock \showarticletitle{AMD 3DNow! Technology: Architecture and
  Implementations}.
\newblock \bibinfo{journal}{\emph{IEEE Micro}} \bibinfo{volume}{19},
  \bibinfo{number}{2} (\bibinfo{date}{March} \bibinfo{year}{1999}),
  \bibinfo{pages}{37--48}.
\newblock
\showISSN{0272-1732}
\urldef\tempurl%
\url{https://doi.org/10.1109/40.755466}
\showDOI{\tempurl}


\bibitem[\protect\citeauthoryear{Pokam, Bihan, Simonnet, and Bodin}{Pokam
  et~al\mbox{.}}{2004}]%
        {swarp}
\bibfield{author}{\bibinfo{person}{Gilles Pokam}, \bibinfo{person}{St{\'e}phane
  Bihan}, \bibinfo{person}{Julien Simonnet}, {and}
  \bibinfo{person}{Fran{\c{c}}ois Bodin}.} \bibinfo{year}{2004}\natexlab{}.
\newblock \showarticletitle{SWARP: a retargetable preprocessor for multimedia
  instructions}.
\newblock \bibinfo{journal}{\emph{Concurrency and Computation: Practice and
  Experience}} \bibinfo{volume}{16}, \bibinfo{number}{2-3}
  (\bibinfo{year}{2004}), \bibinfo{pages}{303--318}.
\newblock


\bibitem[\protect\citeauthoryear{Puschel, Moura, Johnson, Padua, Veloso,
  Singer, Xiong, Franchetti, Gacic, Voronenko, Chen, Johnson, and
  Rizzolo}{Puschel et~al\mbox{.}}{2005}]%
        {spiraln}
\bibfield{author}{\bibinfo{person}{M. Puschel}, \bibinfo{person}{J.~M.~F.
  Moura}, \bibinfo{person}{J.~R. Johnson}, \bibinfo{person}{D. Padua},
  \bibinfo{person}{M.~M. Veloso}, \bibinfo{person}{B.~W. Singer},
  \bibinfo{person}{Jianxin Xiong}, \bibinfo{person}{F. Franchetti},
  \bibinfo{person}{A. Gacic}, \bibinfo{person}{Y. Voronenko},
  \bibinfo{person}{K. Chen}, \bibinfo{person}{R.~W. Johnson}, {and}
  \bibinfo{person}{N. Rizzolo}.} \bibinfo{year}{2005}\natexlab{}.
\newblock \showarticletitle{SPIRAL: Code Generation for DSP Transforms}.
\newblock \bibinfo{journal}{\emph{Proc. IEEE}} \bibinfo{volume}{93},
  \bibinfo{number}{2} (\bibinfo{date}{Feb} \bibinfo{year}{2005}),
  \bibinfo{pages}{232--275}.
\newblock
\showISSN{0018-9219}
\urldef\tempurl%
\url{https://doi.org/10.1109/JPROC.2004.840306}
\showDOI{\tempurl}


\bibitem[\protect\citeauthoryear{Rosen, Wegman, and Zadeck}{Rosen
  et~al\mbox{.}}{1988}]%
        {ssa-form}
\bibfield{author}{\bibinfo{person}{B.~K. Rosen}, \bibinfo{person}{M.~N.
  Wegman}, {and} \bibinfo{person}{F.~K. Zadeck}.}
  \bibinfo{year}{1988}\natexlab{}.
\newblock \showarticletitle{Global Value Numbers and Redundant Computations}.
  In \bibinfo{booktitle}{\emph{Proceedings of the 15th ACM SIGPLAN-SIGACT
  Symposium on Principles of Programming Languages}}
  \emph{(\bibinfo{series}{POPL '88})}. \bibinfo{publisher}{ACM},
  \bibinfo{address}{New York, NY, USA}, \bibinfo{pages}{12--27}.
\newblock
\showISBNx{0-89791-252-7}
\urldef\tempurl%
\url{https://doi.org/10.1145/73560.73562}
\showDOI{\tempurl}


\bibitem[\protect\citeauthoryear{Rosen, Nuzman, and Zaks}{Rosen
  et~al\mbox{.}}{2007}]%
        {Rosen2007}
\bibfield{author}{\bibinfo{person}{Ira Rosen}, \bibinfo{person}{Dorit Nuzman},
  {and} \bibinfo{person}{Ayal Zaks}.} \bibinfo{year}{2007}\natexlab{}.
\newblock \showarticletitle{{Loop-Aware SLP in GCC}}. In
  \bibinfo{booktitle}{\emph{Proceedings of the GCC Developers' Summit}}.
  \bibinfo{pages}{131--142}.
\newblock


\bibitem[\protect\citeauthoryear{Sreraman and Govindarajan}{Sreraman and
  Govindarajan}{2000}]%
        {vlp01}
\bibfield{author}{\bibinfo{person}{N. Sreraman} {and} \bibinfo{person}{R.
  Govindarajan}.} \bibinfo{year}{2000}\natexlab{}.
\newblock \showarticletitle{A Vectorizing Compiler for Multimedia Extensions}.
\newblock \bibinfo{journal}{\emph{Int. J. Parallel Program.}}
  \bibinfo{volume}{28}, \bibinfo{number}{4} (\bibinfo{date}{Aug.}
  \bibinfo{year}{2000}), \bibinfo{pages}{363--400}.
\newblock
\showISSN{0885-7458}
\urldef\tempurl%
\url{https://doi.org/10.1023/A:1007559022013}
\showDOI{\tempurl}


\bibitem[\protect\citeauthoryear{Wang, Wu, Tanase, Serrano, and Moreira}{Wang
  et~al\mbox{.}}{2014}]%
        {generic-simd}
\bibfield{author}{\bibinfo{person}{Haichuan Wang}, \bibinfo{person}{Peng Wu},
  \bibinfo{person}{Ilie~Gabriel Tanase}, \bibinfo{person}{Mauricio~J Serrano},
  {and} \bibinfo{person}{Jos{\'e}~E Moreira}.} \bibinfo{year}{2014}\natexlab{}.
\newblock \showarticletitle{Simple, portable and fast SIMD intrinsic
  programming: generic simd library}. In \bibinfo{booktitle}{\emph{Proceedings
  of the 2014 Workshop on Programming models for SIMD/Vector processing}}. ACM,
  \bibinfo{pages}{9--16}.
\newblock


\bibitem[\protect\citeauthoryear{Willhalm, Oukid, M{\"u}ller, and
  Faerber}{Willhalm et~al\mbox{.}}{2013}]%
        {willhalm2013vectorizing}
\bibfield{author}{\bibinfo{person}{Thomas Willhalm}, \bibinfo{person}{Ismail
  Oukid}, \bibinfo{person}{Ingo M{\"u}ller}, {and} \bibinfo{person}{Franz
  Faerber}.} \bibinfo{year}{2013}\natexlab{}.
\newblock \showarticletitle{Vectorizing Database Column Scans with Complex
  Predicates.}. In \bibinfo{booktitle}{\emph{Fourth International Workshop on
  Accelerating Data Management Systems Using Modern Processor and Storage
  Architectures (ADMS) at VLDB}}.
\newblock


\bibitem[\protect\citeauthoryear{Willhalm, Popovici, Boshmaf, Plattner, Zeier,
  and Schaffner}{Willhalm et~al\mbox{.}}{2009}]%
        {willhalm2009simd}
\bibfield{author}{\bibinfo{person}{Thomas Willhalm}, \bibinfo{person}{Nicolae
  Popovici}, \bibinfo{person}{Yazan Boshmaf}, \bibinfo{person}{Hasso Plattner},
  \bibinfo{person}{Alexander Zeier}, {and} \bibinfo{person}{Jan Schaffner}.}
  \bibinfo{year}{2009}\natexlab{}.
\newblock \showarticletitle{SIMD-scan: ultra fast in-memory table scan using
  on-chip vector processing units}.
\newblock \bibinfo{journal}{\emph{Proceedings of the VLDB Endowment}}
  \bibinfo{volume}{2}, \bibinfo{number}{1}, \bibinfo{pages}{385--394}.
\newblock


\bibitem[\protect\citeauthoryear{Zhou and Xue}{Zhou and Xue}{2016}]%
        {dataReorg}
\bibfield{author}{\bibinfo{person}{Hao Zhou} {and} \bibinfo{person}{Jingling
  Xue}.} \bibinfo{year}{2016}\natexlab{}.
\newblock \showarticletitle{Exploiting Mixed SIMD Parallelism by Reducing Data
  Reorganization Overhead}. In \bibinfo{booktitle}{\emph{Proceedings of the
  2016 International Symposium on Code Generation and Optimization}}
  \emph{(\bibinfo{series}{CGO '16})}. \bibinfo{publisher}{ACM},
  \bibinfo{address}{New York, NY, USA}, \bibinfo{pages}{59--69}.
\newblock
\showISBNx{978-1-4503-3778-6}
\urldef\tempurl%
\url{https://doi.org/10.1145/2854038.2854054}
\showDOI{\tempurl}


\end{thebibliography}

\end{document}